\pdfoutput=1
\documentclass[longauth]{aa} 
\usepackage{graphicx}
\usepackage{epstopdf}

\usepackage{url,hyperref}

\usepackage{txfonts}
\usepackage{amssymb, amsmath, amsfonts} 
\usepackage{bm} 

\newcommand{\g}{\ensuremath{\gamma}} 
\newcommand{\gr}{\mbox{\ensuremath{\g}-ray}}
\newcommand{\xr}{\mbox{X-ray}}
\newcommand{\xrs}{\mbox{X-rays}}
\newcommand{\grs}{\mbox{\ensuremath{\g}-rays}}
\newcommand{\HESS}{H.E.S.S.}
\newcommand{\XMM}{\emph{XMM-Newton}}
\newcommand{\SUZAKU}{\emph{Suzaku}}
\newcommand{\FERMI}{\emph{Fermi}-LAT}
\newcommand{\PARKES}{Parkes}
\newcommand{\ROSAT}{\emph{ROSAT}}
\newcommand{\ASCA}{\emph{ASCA}}
\newcommand{\velajr}{RX\,J0852.0$-$4622}

\newcommand{\psr}{PSR\,J0855$-$4644}
\newcommand{\pmn}{PMN\,J0852$-$4712}
\newcommand{\twofgl}{2FGL\,J0853.5$-$4711}
\newcommand{\threefgl}{3FGL\,J0852.7$-$4631e}
\newcommand{\rxj}{RX\,J1713.7$-$3946}
\newcommand{\UNITS}[1]{\ensuremath{\,\mathrm{#1}}}
\newcommand{\UNITSwoS}[1]{\ensuremath{\mathrm{#1}}} 
\newcommand{\dg}{\ensuremath{^{\circ}}} 
\newcommand{\on}{ON}
\newcommand{\off}{OFF}
\newcommand{\Non}{\ensuremath{N_\mathrm{\on}}}
\newcommand{\Noff}{\ensuremath{N_\mathrm{\off}}}
\newcommand{\stat}{\ensuremath{_{\textrm{\tiny{stat}}}}}
\newcommand{\syst}{\ensuremath{_{\textrm{\tiny{syst}}}}}
\newcommand{\lik}{\ensuremath{L}} 

\newcounter{mylabelcounter}
\makeatletter
\newcommand{\labelText}[2]{%
#1\refstepcounter{mylabelcounter}%
\immediate\write\@auxout{%
  \string\newlabel{#2}{{1}{\thepage}{{\unexpanded{#1}}}{mylabelcounter.\number\value{mylabelcounter}}{}}%
}%
}
\makeatother

\titlerunning{\velajr: Morphology studies and resolved spectroscopy}
\authorrunning{The \HESS\ Collaboration}

\begin{document}

   \title{Deeper \HESS\ observations of Vela Junior (\velajr):
          \\Morphology studies and resolved spectroscopy\nameref{text:footnotemark_title}} 


\author{\small H.E.S.S. Collaboration
\and H.~Abdalla \inst{1}
\and A.~Abramowski \inst{2}
\and F.~Aharonian \inst{3,4,5}
\and F.~Ait Benkhali \inst{3}
\and A.G.~Akhperjanian\protect\nameref{text:footnotemark_deceased} \inst{6,5} 
\and T.~Andersson \inst{10}
\and E.O.~Ang\"uner \inst{21}
\and M.~Arakawa \inst{43}
\and M.~Arrieta \inst{15}
% ...
}

\institute{
Centre for Space Research, North-West University, Potchefstroom 2520, South Africa \and
Universit\"at Hamburg, Institut f\"ur Experimentalphysik, Luruper Chaussee 149, D 22761 Hamburg, Germany \and
Max-Planck-Institut f\"ur Kernphysik, P.O. Box 103980, D 69029 Heidelberg, Germany \and
Dublin Institute for Advanced Studies, 31 Fitzwilliam Place, Dublin 2, Ireland \and
National Academy of Sciences of the Republic of Armenia, Marshall Baghramian Avenue, 24, 0019 Yerevan, Republic of Armenia \and
Yerevan Physics Institute, 2 Alikhanian Brothers St., 375036 Yerevan, Armenia \and
Institut f\"ur Physik, Humboldt-Universit\"at zu Berlin, Newtonstr. 15, D 12489 Berlin, Germany \and
University of Namibia, Department of Physics, Private Bag 13301, Windhoek, Namibia \and
GRAPPA, Anton Pannekoek Institute for Astronomy, University of Amsterdam, Science Park 904, 1098 XH Amsterdam, The Netherlands \and
Department of Physics and Electrical Engineering, Linnaeus University, 351 95 V\"axj\"o, Sweden \and
Institut f\"ur Theoretische Physik, Lehrstuhl IV: Weltraum und Astrophysik, Ruhr-Universit\"at Bochum, D 44780 Bochum, Germany \and
GRAPPA, Anton Pannekoek Institute for Astronomy and Institute of High-Energy Physics, University of Amsterdam, Science Park 904, 1098 XH Amsterdam, The Netherlands \and
Institut f\"ur Astro- und Teilchenphysik, Leopold-Franzens-Universit\"at Innsbruck, A-6020 Innsbruck, Austria \and
School of Physical Sciences, University of Adelaide, Adelaide 5005, Australia \and
LUTH, Observatoire de Paris, PSL Research University, CNRS, Universit\'e Paris Diderot, 5 Place Jules Janssen, 92190 Meudon, France \and
Sorbonne Universit\'es, UPMC Universit\'e Paris 06, Universit\'e Paris Diderot, Sorbonne Paris Cit\'e, CNRS, Laboratoire de Physique Nucl\'eaire et de Hautes Energies (LPNHE), 4 place Jussieu, F-75252, Paris Cedex 5, France \and
Laboratoire Univers et Particules de Montpellier, Universit\'e Montpellier, CNRS/IN2P3, CC 72, Place Eug\`ene Bataillon, F-34095 Montpellier Cedex 5, France \and
DSM/Irfu, CEA Saclay, F-91191 Gif-Sur-Yvette Cedex, France \and
Astronomical Observatory, The University of Warsaw, Al. Ujazdowskie 4, 00-478 Warsaw, Poland \and
Aix Marseille Universit\'e, CNRS/IN2P3, CPPM UMR 7346, 13288 Marseille, France \and
Instytut Fizyki J\c{a}drowej PAN, ul. Radzikowskiego 152, 31-342 Krak{\'o}w, Poland \and
Funded by EU FP7 Marie Curie, grant agreement No. PIEF-GA-2012-332350, \and
School of Physics, University of the Witwatersrand, 1 Jan Smuts Avenue, Braamfontein, Johannesburg, 2050 South Africa \and
Laboratoire d'Annecy-le-Vieux de Physique des Particules, Universit\'{e} Savoie Mont-Blanc, CNRS/IN2P3, F-74941 Annecy-le-Vieux, France \and
Landessternwarte, Universit\"at Heidelberg, K\"onigstuhl, D 69117 Heidelberg, Germany \and
Universit\'e Bordeaux, CNRS/IN2P3, Centre d'\'Etudes Nucl\'eaires de Bordeaux Gradignan, 33175 Gradignan, France \and
Oskar Klein Centre, Department of Physics, Stockholm University, Albanova University Center, SE-10691 Stockholm, Sweden \and
Wallenberg Academy Fellow, \and
Institut f\"ur Astronomie und Astrophysik, Universit\"at T\"ubingen, Sand 1, D 72076 T\"ubingen, Germany \and
Laboratoire Leprince-Ringuet, Ecole Polytechnique, CNRS/IN2P3, F-91128 Palaiseau, France \and
APC, AstroParticule et Cosmologie, Universit\'{e} Paris Diderot, CNRS/IN2P3, CEA/Irfu, Observatoire de Paris, Sorbonne Paris Cit\'{e}, 10, rue Alice Domon et L\'{e}onie Duquet, 75205 Paris Cedex 13, France \and
Univ. Grenoble Alpes, IPAG, F-38000 Grenoble, France \protect\\ CNRS, IPAG, F-38000 Grenoble, France \and
Department of Physics and Astronomy, The University of Leicester, University Road, Leicester, LE1 7RH, United Kingdom \and
Nicolaus Copernicus Astronomical Center, Polish Academy of Sciences, ul. Bartycka 18, 00-716 Warsaw, Poland \and
Institut f\"ur Physik und Astronomie, Universit\"at Potsdam, Karl-Liebknecht-Strasse 24/25, D 14476 Potsdam, Germany \and
Friedrich-Alexander-Universit\"at Erlangen-N\"urnberg, Erlangen Centre for Astroparticle Physics, Erwin-Rommel-Str. 1, D 91058 Erlangen, Germany \and
DESY, D-15738 Zeuthen, Germany \and
Obserwatorium Astronomiczne, Uniwersytet Jagiello{\'n}ski, ul. Orla 171, 30-244 Krak{\'o}w, Poland \and
Centre for Astronomy, Faculty of Physics, Astronomy and Informatics, Nicolaus Copernicus University, Grudziadzka 5, 87-100 Torun, Poland \and
Department of Physics, University of the Free State, PO Box 339, Bloemfontein 9300, South Africa \and
Heisenberg Fellow (DFG), ITA Universit\"at Heidelberg, Germany \and
GRAPPA, Institute of High-Energy Physics, University of Amsterdam, Science Park 904, 1098 XH Amsterdam, The Netherlands \and
Department of Physics, Rikkyo University, 3-34-1 Nishi-Ikebukuro, Toshima-ku, Tokyo 171-8501, Japan
% ...
}

\offprints{H.E.S.S.~Collaboration,
\protect\\\email{\href{mailto:contact.hess@hess-experiment.eu}{contact.hess@hess-experiment.eu}};
\protect\\\protect\labelText{$^\bigstar$}{text:footnotemark_title} FITS image and spectrum tables available online at CDS at
\protect\\\url{http://cdsarc.u-strasbg.fr/viz-bin/qcat?J/A+A/612/A7} 
\protect\\\protect\labelText{$^\dagger$}{text:footnotemark_deceased} Deceased
\protect\\\protect\labelText{$^{\bigstar\bigstar}$}{text:footnotemark_ca} Corresponding authors
}

\date{Published on 9 April 2018 as: H.E.S.S. Collaboration, A\&A 612, A7 (2018)}


  \abstract
   {}
   {
    We study \gr\ emission from the shell-type supernova remnant (SNR) \velajr\ to better characterize its spectral properties and its distribution over the SNR.
   }
   {
    The analysis of an extended High Energy Spectroscopic System (\HESS) data set at very high energies ($E>100$\,GeV) permits detailed studies, as well as spatially
resolved spectroscopy, of the morphology and spectrum of the whole \velajr\ region.
    The \HESS\ data are combined with archival data from other wavebands and interpreted in the framework of leptonic and hadronic models.
    The joint \FERMI-\HESS\ spectrum allows the direct determination of the spectral characteristics of the parent particle population in leptonic and hadronic scenarios using only GeV-TeV data.
   }
   {
    An updated analysis of the \HESS\ data shows that the spectrum of the entire SNR connects smoothly to the high-energy spectrum measured by \FERMI.
    The increased data set makes it possible to demonstrate that the \HESS\
    spectrum deviates significantly from a power law and is well described by both a curved power law and a power law with an exponential
    cutoff at an energy of $E_\mathrm{cut} = (6.7 \pm 1.2\stat \pm 1.2\syst) \UNITS{TeV}$.
    The joint \FERMI-\HESS\ spectrum allows the unambiguous identification of the spectral shape as a power law with an exponential cutoff.
    No significant evidence is found for a variation of the spectral parameters across the SNR, suggesting similar conditions of particle acceleration across the remnant.
    A simple modeling using one particle population to model the SNR emission demonstrates that both leptonic and hadronic emission scenarios remain plausible.
    It is also shown that at least a part of the shell emission is likely due to the presence of a pulsar wind nebula around \psr.
   }
   {}


   \maketitle

%

\section{Introduction}\label{sec:intro}

\object{RX J0852.0$-$4622} belongs to the class of young shell-type supernova remnants (SNRs) that display broadband nonthermal
emission and have been detected at very high energies \citep[photon energies $E>100$\,GeV;][]{paper:vjr_cangaroo, paper:vjr_hess_paper1, paper:vjr_hess_paper2}; this SNR is listed in the Green SNR catalog
\citep{cat:green_snr} as \object{SNR G266.2$-$1.2}, as \object{HESS J0852$-$463}
in the High Energy Spectroscopic System (\HESS) catalog\footnote{https://www.mpi-hd.mpg.de/hfm/HESS/pages/home/sources/},
and is commonly referred to as \emph{\object{Vela Junior}}.
Its properties are similar to \rxj\ \citep{RXJ1713Forth} and, similar to this object, it has been extensively
studied at multiple wavelengths \citep[for an overview see][]{paper:vjr_hess_paper2}. The distance to the remnant and its age are still
under debate in the literature but the range of possible values is narrowing. Studies of the shell expansion in \xrs, based on \XMM\ data and the assumption of a shock velocity of $3000 \UNITS{km\, s^{-1}}$, have established a distance of $\sim\!\!750$\,pc and an age between 1700 yr and 4300 yr \citep{2008ApJ...678L..35K}.
Similar studies were also recently conducted with \emph{Chandra} data from the years 2003 to 2008, placing a lower limit on the distance to the remnant
at $500$\,pc \citep{2015ApJ...798...82A}. An upper limit is determined by the comparison of the detected $^{44}\mbox{Ti}$ line emission with
evolution models for different types of supernova (SN) explosions, yielding a highest allowed distance of $1$\,kpc \citep{2010A&A...519A..86I}.
The distance estimate by \citet{2008ApJ...678L..35K} is in the middle of the range of allowed values and is adopted hereafter.

The \gr\ emission from \velajr\ has been interpreted in the framework of both hadronic
(proton-proton interactions with subsequent $\pi^0$ decay) and leptonic (inverse Compton (IC)
scattering of relativistic electrons on ambient radiation fields) scenarios, without a definite answer so far.
The high magnetic fields implied by hadronic scenarios are supported by the existence of sharp filamentary \xr\ structures in the northwestern (NW) rim of the remnant,
which have been resolved by \emph{Chandra} \citep{2005ApJ...632..294B}. The small effective width of these structures is explained by fast synchrotron
cooling of relativistic electrons implying strong magnetic field amplification. The estimates of the amplified magnetic field strength range from $\gtrsim\!\!100\,\mu$G \citep{2009A&A...505..641B} to $\sim\!\!500\,\mu$G \citep{2005ApJ...632..294B}. A softening of the \xr\ spectrum of the remnant
toward the interior of the SNR recently detected in the NW rim of \velajr\ with \XMM\ was interpreted as the gradual
decrease of the cutoff energy of the electron spectrum due to fast synchrotron cooling \citep{2013A&A...551A.132K}.
The authors have shown that the detected softening can be reproduced for a low magnetic field of a few $\mu$G, which would, however,
require a high value of the electron cutoff energy of $\sim\!\!100$\,TeV near the shock front.
For a lower value of the cutoff energy of about 20\,TeV, a high magnetic field of $\gtrsim\!\!100\,\mu$G is required,
still suggesting a strong magnetic field amplification in the rim.
Alternatively, the sharp X-ray filaments
can be explained by nonlinear damping of strong magnetic turbulence downstream from the shock, which creates thin regions with an enhanced magnetic
field strength \citep[see, e.g.,][]{2005ApJ...626L.101P}. In this case the width of the filaments is limited by the extent of the regions with strong magnetic
fields and the softening of the spectrum can be explained by the spatial variation of the magnetic field strength
\citep{2005ApJ...626L.101P, 2012A&A...545A..47R}.

On the other hand, the hadronic scenario
requires a relatively high density of the ambient medium of about $\sim\!\!1\,\mathrm{cm}^{-3}$ \citep{paper:vjr_hess_paper2}, which is in contradiction
with the lack of detected thermal \xr\ emission. The lack of thermal emission places an upper limit on the ambient density at $\sim\!\!0.01\,\mathrm{cm}^{-3}$
\citep{2001ApJ...548..814S}. Such a low density of the ambient medium is actually expected in the case of a
core collapse SN explosion when the SNR is expanding inside the stellar
bubble of the progenitor star. This scenario is argued to be the case for \velajr, as supported by the detection of the
central compact object (CCO) AX\,J0851.9$-$4617.4\footnote{Also known as CXOU\,J085201.4$-$461753.} close to the center
of the remnant \citep{1998Natur.396..141A, 1999A&A...350..997A, 2001ApJ...548..814S}. However, no pulsations were
detected from the CCO and, moreover, later it was suggested that the source might be an unrelated planetary nebula
\citep{2006A&A...449..243R}. An ambient density as low as $0.01\,\mathrm{cm}^{-3}$ leads to an unrealistically
high value of the total energy in protons of $\sim\!\!10^{51}$\,erg in the hadronic scenario \citep{paper:vjr_fermi}.
This would imply that all the energy from the SN explosion ($\sim\!\!10^{51}$\,erg) is used in accelerating particles,
whereas normally this number is expected to be a factor $\sim\!\!10$ lower. Several explanations were suggested to overcome the
problem of the inconsistency of the hadronic model with the lack of thermal \xr\ emission. \citet{2009APh....31..431T} suggested
that \velajr\ might be older ($17500$\,yr) than usually assumed, which would place it in the transition phase between
the adiabatic and radiative phases of SNR evolution. In this case the expected thermal \xr\ flux decreases due to the
formation of the dense shell and the cooling of the gas. The lack of thermal emission can only be
claimed for the gas with a temperature above 1\,keV because at lower energies \velajr\ is completely obscured
by the strong thermal emission from the Vela\,SNR. Another possible explanation was suggested by \citet{2012ApJ...744...71I} and later by \citet{2014MNRAS.445L..70G} for
the \rxj\ SNR showing that a highly inhomogeneous clumpy environment significantly increases the expected \gr\
emission from hadronic interactions, while the thermal \xr\ emission might remain at a very low level because denser cores of gas
in clumps, which carry most of the mass, can survive the shock passing without being ionized. The environment of \velajr\ exhibits
a number of large clouds of atomic hydrogen, which are close to the current position of the main shock
and coincident with the regions of enhanced emission from the remnant \citep[see, e.g.,][and references therein]{2013ASSP...34..249F, 2014MNRAS.437..976O}.
The lack of thermal \xr\ emission is at the same time a strong argument in favor of a leptonic scenario.
The leptonic scenario can naturally explain the correlation between the \xr\ and \gr\ emitting regions \citep{paper:vjr_hess_paper2} and
is able to match the broadband emission from \velajr. This\ suggests, however, a very low average magnetic
field of about $5\!\!-\!\!10\,\mu$G \citep{2013ApJ...767...20L, paper:vjr_hess_paper2}, which, in turn, seems to be in contradiction with the sharp filamentary structures detected
in \xrs. Recent \xr\ observations with \SUZAKU\ \citep{2016PASJ...68S..10T} reveal a faint hard \xr\ component in the NW rim of \velajr. \xrs\ are reported with a spectral index of $3.15^{+1.18}_{-1.14}$ in the energy range from 12 keV to 22 keV.
The absence of roll-off in the \xr\ spectrum disfavors one-zone synchrotron models with electron spectra in the form
of a power law with an exponential cutoff. However, the uncertainties on the spectral parameters are too large to draw strong conclusions.

\xr\ observations with \XMM\ \citep{2013A&A...551A...7A} led to the discovery of a pulsar wind nebula (PWN) with an extension of 150\,arcsec
around the energetic (spin-down power $\dot{E} = 1.1 \times 10^{36}$\,erg/s) radio pulsar \object{PSR J0855$-$4644} that coincides with the shell of \velajr.
The pulsar is energetic enough to power a very high-energy PWN that is detectable by current generation Cherenkov telescopes \citep[and references therein]{paper:hess_pwnpop}.
Measurements of the column density
toward the pulsar and \velajr\ in \xrs\ with \XMM\ data show that both objects are located at similar distances. Nevertheless,
an association of \psr\ with the SNR seems unlikely to be due to the age difference
(the pulsar characteristic age is $140 \UNITS{kyr}$)
and the large speed ($\sim\!\!3000 \UNITS{km\, s^{-1}}$) needed by the pulsar to travel from the geometrical center of the SNR to its current position.

This paper reports new \HESS\ observations of \velajr. Section~\ref{sec:obs_and_analysis} describes
the data set and the applied analysis techniques, and Sect.~\ref{sec:results} presents the results
of the conducted morphological and spectral studies.
The results are discussed in the context
of multiwavelength data and a simple modeling using one particle population for the entire SNR emission in Sect.~\ref{sec:interpretation}.
%

\section{\HESS\ observations and data analysis}\label{sec:obs_and_analysis}

\HESS\ is an array of five Cherenkov telescopes situated in the Khomas Highland of Namibia at an altitude of 1800\,m above sea level.
In its initial phase, it consisted of four 13\,m diameter telescopes sensitive in
the energy range of 100\,GeV to 100\,TeV. In 2012, a fifth, 28\,m telescope was added at the center of the array that
allows the threshold of the instrument to be lowered to several tens of GeV. All data presented in this work
were taken in the initial phase of \HESS\ when only the 13\,m telescopes were available. More details on the performance of \HESS\
in the four-telescope configuration are given in \citet{paper:hess_crab} and references therein.

The data used for the analysis of \velajr\ were taken between 2004 and 2009; the bulk of the observations were performed between
2004 and 2006. The analysis of the data up to the end of 2005 has already been published \citep{paper:vjr_hess_paper1,paper:vjr_hess_paper2}.
This work presents a reanalysis of the source motivated by a rough doubling of the data set.
The exposure times (normalized to an offset of $0.7 \UNITSwoS{\dg}$) available for morphological and spectral analyses amount to 60\,h and 39\,h, respectively.
Thanks to the improved statistics more detailed studies of the morphology and spectrum are possible. In addition, a spatially resolved spectroscopy of the source can be performed.

The \velajr\ data
were analyzed with methods that are similar to the techniques discussed in \citet{paper:hess_crab},
however with two differences: gamma-hadron separation took advantage of a multivariate approach
\citep[][instead of box cuts based on scaled Hillas parameters]{paper:hess_tmva},
and a forward-folding method with a likelihood technique similar to that described in \citet{paper:ff_spectrum_piron}
was used for spectrum derivation and flux measurements (instead of least-squares fitting). The results presented here
were derived with the so-called HAP-HD analysis software
and the corresponding instrument response tables.
All results have been cross-checked with
two independent software chains using an independent calibration of the data, yielding compatible results.

The extraction region or \on-source region (short: \on\ region) selected for the analysis is a circle centered at the nominal
position of \velajr\ (in right ascension and declination 8h52m, $-46\dg22'12''$, J2000) with a radius of 1.0\dg. This \on\ region
is referred to as \emph{whole SNR} in the following. To avoid any contamination from the nearby Vela X PWN \citep{velax} the region of the sky
covering this source is excluded from the analysis, in particular in the estimation of cosmic-ray backgrounds from \off-source regions (short: \off\ regions).

Two different analysis configurations were applied to the data when studying the morphological and spectral properties of \velajr, respectively:
\begin{itemize}
  \item \emph{Spatial} analysis was used to produce two-dimensional spatial skymaps to study the morphology of \velajr.
        This analysis is characterized by a loose (less stringent) data selection criterion in which data taken under optimal atmospherical conditions, but varying
        instrumental conditions, are acceptable.
        Hard image cuts were used for data filtering to improve the angular resolution at the expense of statistics.
        Cosmic-ray backgrounds were estimated with the \emph{ring background model} \citep{paper:bg}.
  \item \emph{Spectral} analysis was used to produce flux measurements to study the photon spectrum of the entire
        SNR or of subregions (i.e.,~for spatially resolved spectroscopy). This analysis is characterized by
        a stringent (conservative) data selection criterion aiming at reducing systematic errors by selecting only data taken under optimal atmospherical and instrumental conditions.
        Standard cuts were used for data filtering, and the \emph{reflected region background model} \citep{paper:bg} was
        applied for cosmic-ray background estimation.
\end{itemize}
Details of the data selection criteria and the cut configurations (hard, standard) can be found in \citet{paper:hess_crab}.
The analysis of \velajr\ is challenging since the regions used for signal extraction and for excluding nearby sources from the analysis are large with respect to the field of view of the instrument, as compared to other sources.
A careful study of the systematic uncertainties showed a larger than usual variation of
spectral parameters when comparing the results derived with different analysis chains in use within the \HESS\ Collaboration. The dispersion of the results yielded a systematic error on flux measurements of 25\% and an error on spectral indices
of 0.2 for spectral indices in the range from 1.5 to 2.2. The error of cutoff energies was found to be 20\%.
These uncertainties are slightly larger than the typical systematic errors given for \HESS\ in \citet{paper:hess_crab},
and do not represent the general trend of \HESS\ measurements.
%

\section{Results}\label{sec:results}

\begin{table*}[!htb]
  \centering
  \begin{tabular}{llcccccccccc}
  \hline
  region & analysis & $\langle\theta_\mathrm{zen}\rangle$ & $\langle\theta_\mathrm{az}\rangle$ & $\langle\theta_\mathrm{off}\rangle$ & $t\,[\UNITSwoS{h}]$ & \Non & \Noff & $\alpha$ & excess & significance \\
  \hline
  \hline
  whole SNR & spatial  & 29\UNITSwoS{\dg} & 194\UNITSwoS{\dg} & 1.4\UNITSwoS{\dg}  & 93.6 & 34025 & 16854 & 1.4   & 10332 & $39.1\sigma$  \\
  whole SNR & spectral & 32\UNITSwoS{\dg} & 207\UNITSwoS{\dg} & 1.2\UNITSwoS{\dg}  & 21.0 & 43363 & 36097 & 1.0   &  7266 & $25.8\sigma$  \\
  \hline
  NW rim    & spectral & 32\UNITSwoS{\dg} & 204\UNITSwoS{\dg} & 1.4\UNITSwoS{\dg}  & 28.8 & 17561 & 32212 & 0.4   &  4232 & $29.0\sigma$  \\
  \hline
  0         & spectral & 32\UNITSwoS{\dg} & 206\UNITSwoS{\dg} & 1.2\UNITSwoS{\dg}  & 31.3 & 24873 & 51436 & 0.4   &  3857 & $21.6\sigma$  \\
  1         & spectral & 30\UNITSwoS{\dg} & 201\UNITSwoS{\dg} & 0.95\UNITSwoS{\dg} & 18.4 &  5290 & 18586 & 0.2   &   949 & $12.5\sigma$  \\
  2         & spectral & 31\UNITSwoS{\dg} & 200\UNITSwoS{\dg} & 1.0\UNITSwoS{\dg}  & 27.8 &  6912 & 27412 & 0.2   &  1058 & $12.1\sigma$  \\
  3         & spectral & 32\UNITSwoS{\dg} & 200\UNITSwoS{\dg} & 0.92\UNITSwoS{\dg} & 22.2 &  6639 & 21000 & 0.3   &  1342 & $15.7\sigma$  \\
  4         & spectral & 32\UNITSwoS{\dg} & 202\UNITSwoS{\dg} & 0.91\UNITSwoS{\dg} & 19.1 &  5315 & 12941 & 0.3   &  1062 & $13.5\sigma$  \\
  5         & spectral & 31\UNITSwoS{\dg} & 197\UNITSwoS{\dg} & 1.1\UNITSwoS{\dg}  & 30.8 &  7163 & 27146 & 0.2   &  1469 & $16.9\sigma$  \\
  6         & spectral & 30\UNITSwoS{\dg} & 203\UNITSwoS{\dg} & 0.93\UNITSwoS{\dg} & 17.4 &  5189 & 10461 & 0.4   &  1261 & $16.1\sigma$  \\
  \hline
  A         & spectral & 31\UNITSwoS{\dg} & 200\UNITSwoS{\dg} & 1.2\UNITSwoS{\dg}  & 38.0 &  1151 & 14813 & 0.061 &   248 & $ 7.7\sigma$  \\
  B         & spectral & 31\UNITSwoS{\dg} & 200\UNITSwoS{\dg} & 1.2\UNITSwoS{\dg}  & 38.0 &  8127 & 34927 & 0.19  &  1513 & $16.4\sigma$  \\
  C         & spectral & 32\UNITSwoS{\dg} & 202\UNITSwoS{\dg} & 1.2\UNITSwoS{\dg}  & 35.6 & 27852 & 47530 & 0.49  &  4718 & $24.4\sigma$  \\
  D         & spectral & 32\UNITSwoS{\dg} & 201\UNITSwoS{\dg} & 1.2\UNITSwoS{\dg}  & 38.8 &  1180 & 15587 & 0.057 &   298 & $ 9.3\sigma$  \\
  B$'$      & spectral & 31\UNITSwoS{\dg} & 200\UNITSwoS{\dg} & 1.2\UNITSwoS{\dg}  & 38.0 &  6976 & 29997 & 0.19  &  1264 & $14.7\sigma$  \\
  C$'$      & spectral & 32\UNITSwoS{\dg} & 202\UNITSwoS{\dg} & 1.2\UNITSwoS{\dg}  & 35.6 & 19594 & 33898 & 0.49  &  3144 & $19.3\sigma$  \\
  \hline
  \end{tabular}
  \caption[Analysis statistics]{Statistics of the different analyses of the \velajr\ data set. For each analysis, the table shows the name of the region analyzed (region definitions in Table~\ref{tab:reg_def}), the applied analysis type, the mean zenith, and azimuth angles of the observations ($\langle\theta_\mathrm{zen}\rangle$ and $\langle\theta_\mathrm{az}\rangle$, respectively), the mean offset angle $\langle\theta_\mathrm{off}\rangle$, the livetime $t$, the number of events in the signal (\on) region \Non, the number of events in the background (\off) region \Noff, the exposure normalization (ratio of \on\ to \off\ exposures) $\alpha$, the number of excess counts in the \on\ region, and the significance of the signal in the \on\ region in number of Gaussian standard deviations $\sigma$.}
  \label{tab:stats}
\end{table*}

Table~\ref{tab:stats} summarizes the event statistics of all analyses applied to the data using the \on\ region detailed above to encompass the entire SNR, and other smaller regions, to study spectral details of the different parts of \velajr\ in the spatially resolved spectroscopy studies presented in Sect.~\ref{sec:spec_morph}.
The specific parameters for all regions used are defined in Table~\ref{tab:reg_def}.
%

\subsection{Source morphology}\label{sec:spat_morph}

\begin{figure*}[!htb]
  \centering
    \includegraphics[width=0.45\textwidth]{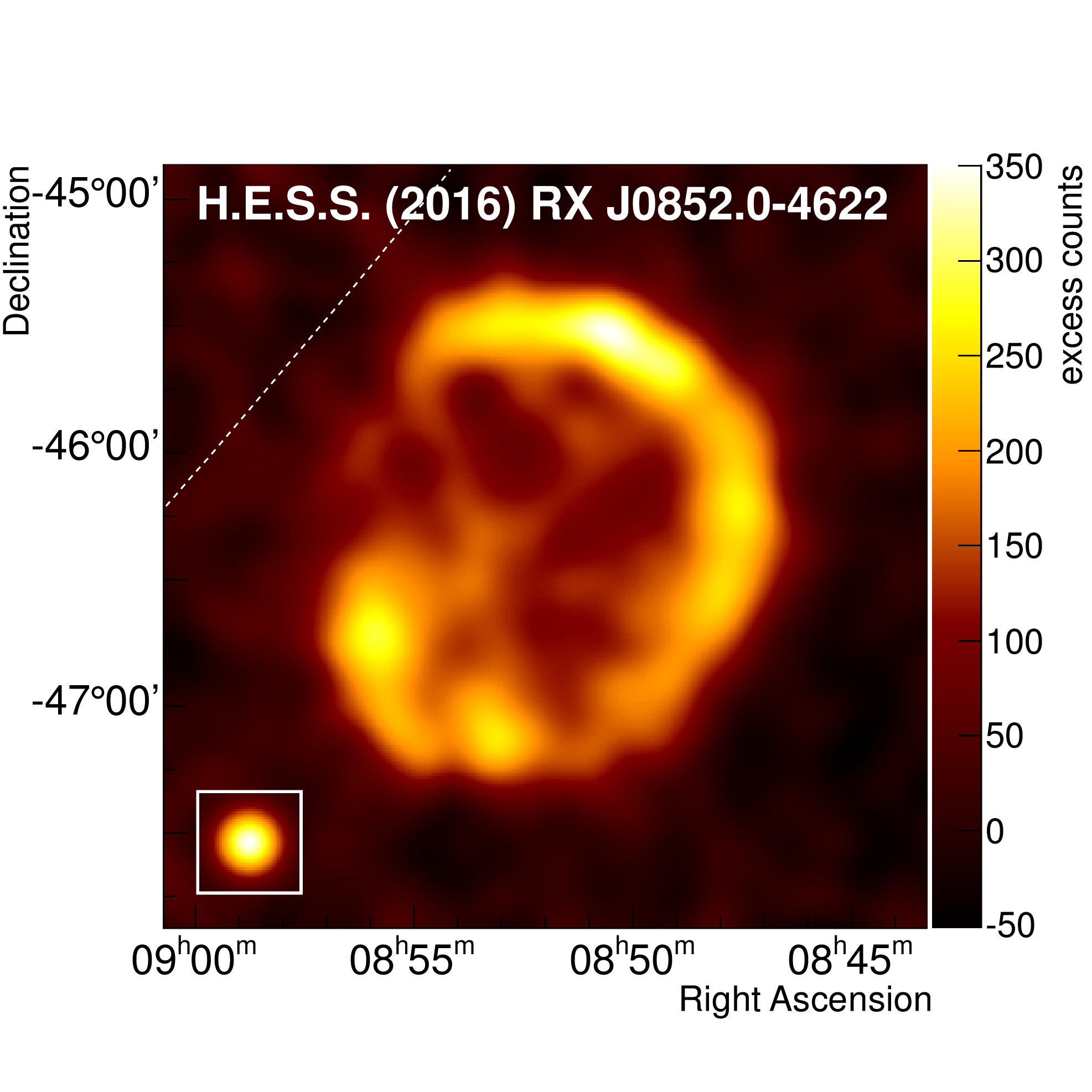}
    \includegraphics[width=0.45\textwidth]{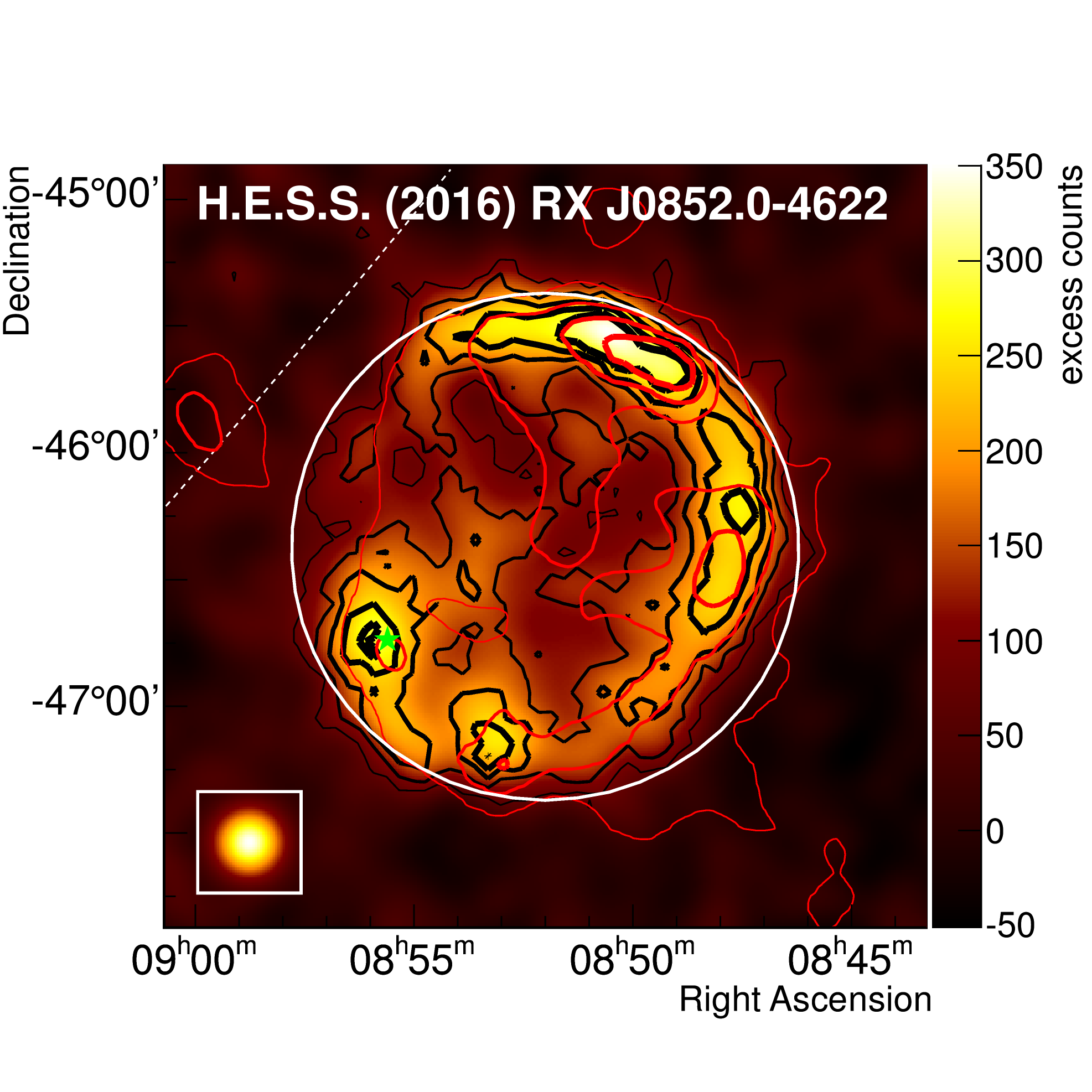}
  \caption[Excess maps]{Left: exposure-corrected excess map for \velajr. The data were binned in bins of 0.01\dg\ on each coordinate and smoothed with a Gaussian function of width 0.08\dg. The white dashed line shows the position of the Galactic plane; the inset shows the PSF of the analysis at the same scale for comparison. Right: same as in the left panel, but additionally the boundary of the \on\ region is shown as a white circle and the significance contours at 3, 5, 7, 9, and $11\sigma$ are shown in black with increasing line width for increasing significance. In addition, \xr\ contours from the \ROSAT\ All Sky Survey for energies larger than 1.3 keV are shown in red. The \xr\ data were smoothed in the same way as the \gr\ data to allow for a direct comparison. The \xr\ contours were derived at 25, 50, 75, and 100 counts. The green star indicates the position of \psr.}
  \label{fig:skymaps}
\end{figure*}

The spatial analysis of \velajr\ (cf.~first row of Table~\ref{tab:stats}) is based on a data set with a total livetime of 93.6\,h
and results in a total significance of 39.1$\sigma$.
The difference between the livetime of the analysis and the exposure time previously quoted is because a large amount of
the available observations were targeting nearby sources, had large offsets, and hence a small effective exposure time.
The angular resolution attained in the morphological analysis is
characterized by a PSF width (68\% containment radius) of 0.08\dg. This angular resolution is larger than the 0.06\dg\ reported for
the morphological analysis in \citet{paper:vjr_hess_paper2} since
the default telescope multiplicity ($\ge\!\!2$ telescopes per event) was respected in the analysis presented here.
This favors \gr\ efficiency at
the expense of a somewhat larger angular resolution. Figure~\ref{fig:skymaps}, left, shows the excess map corrected for the gradient of exposure across the field of view and
smoothed with a Gaussian function with a width equal to the PSF width of 0.08\dg. The TeV emission matches the shape of the SNR
shell well when compared to measurements in radio and \xrs;
the brightest region is a semicircular arc in the NW (the NW rim), and there are two other bright regions toward the south
and southeast.
The contours of the \ROSAT\ All Sky Survey for energies larger than 1.3 keV, smoothed to match the \HESS\ angular resolution, are shown in red in Fig.~\ref{fig:skymaps}, right, together with the \HESS\ significance contours. The overall agreement is good.
With an extension of $\sim\!\!2\dg$ in diameter, \velajr\ is one of the largest known TeV sources in the sky.

The \on\ region (shown in Fig.~\ref{fig:skymaps}, right) is in fact slightly too small since it barely encompasses the $5\sigma$ significance contour.
This region was chosen because of the large extension of the source;
an even larger \on\ region
would make the determination of \off\ regions within the same field of view very difficult, thereby reducing the effective data set available, especially for the spectral studies (cf.~Sect.~\ref{sec:spectrum}).
The number of \grs\ left outside the \on\ region is small ($\sim\!\!5\%$) compared to the signal inside and its systematic
uncertainty. In the following, the effect is therefore neglected.
Nevertheless, to avoid background contamination, a larger region covering the whole \velajr\ emission was excluded for the determination of suitable \off\ regions.

\begin{figure*}[!htb]
  \centering
    \includegraphics[width=0.33\textwidth]{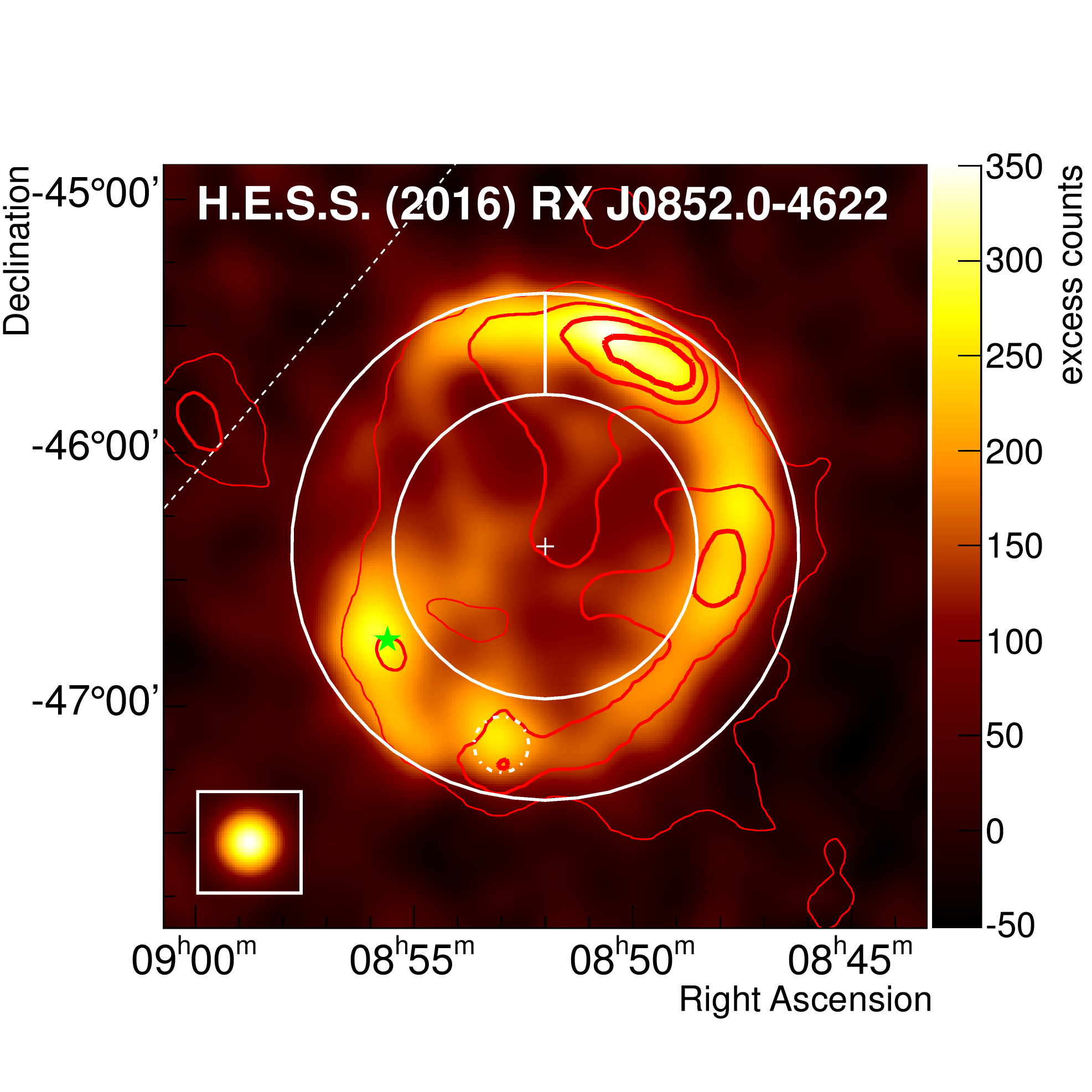}
    \includegraphics[width=0.66\textwidth]{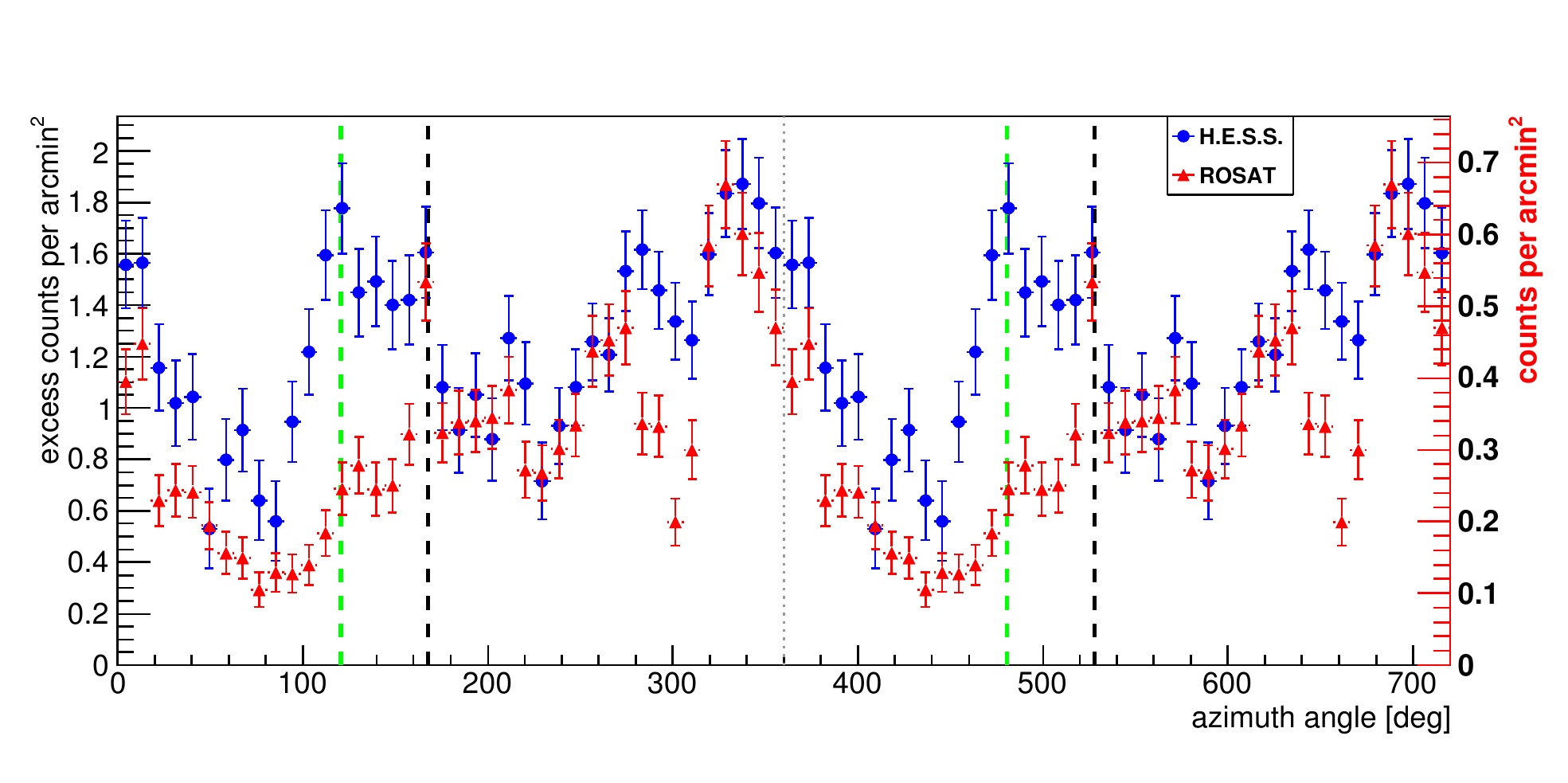}
  \caption[Azimuthal profile]{Left: smoothed and exposure-corrected excess map with \ROSAT\ \xr\ contours, as in Fig.~\ref{fig:skymaps}.
The white annulus denotes the region used for extraction of the azimuthal profile; the white vertical line denotes the origin of the azimuthal angle (north), which increases in the counterclockwise direction. The green star indicates the position of \psr. The white dot-dashed line indicates a point-like region around the position of the southern enhancement. Right: the azimuthal profile extracted from the annulus in the skymap on the left panel is shown in blue for the \HESS\ \gr\ data (left scale) and in red for the \ROSAT\ \xr\ data (right scale). For better visibility, two periods separated by a dashed gray line are shown. The azimuthal position of \psr\ and the center of the region around the southern enhancement are indicated by green and black vertical dashed lines, respectively. The vertical error bars represent $1\sigma$ statistical uncertainties; the horizontal bars represent the bin widths.}
  \label{fig:az_prof}
\end{figure*}

Projections of the skymap in Fig.~\ref{fig:skymaps} were calculated from the unsmoothed data and the resulting photon counts were normalized
to the covered solid angle. Inspection of the radial profile (i.e.,~a skymap projection along the radial coordinate; profile not shown here) confirms the earlier result
that the emission comes from a thin shell and not from a sphere \citep{paper:vjr_hess_paper2}.
The azimuthal profile (i.e.,~skymap projection along the azimuthal coordinate)
calculated for an annulus with inner and outer radius of 0.6\dg\ and 1.0\dg\ , respectively, around the center of \velajr\ is shown in Fig.~\ref{fig:az_prof}.
The azimuth angle is defined counterclockwise from north. Two periods separated by a dashed gray line are shown;
green and black dashed lines at 121\dg\ and 168\dg, respectively, denote the position of \psr\ and the center of the enhancement seen toward the south of the shell.

The azimuthal profile (blue points of Fig.~\ref{fig:az_prof}, right) clearly shows that the emission is not homogeneous along the shell. In general, the NW rim (from $220\UNITSwoS{\dg}$ to $400\UNITSwoS{\dg}$) is brighter than the southeastern part
of the shell, where a region of high emission is observed between the southern ($\sim\!\!160\dg$) and the southeastern ($\sim\!\!120\dg$) enhancements. The emission in the southeast
(cf.~Fig.~\ref{fig:az_prof}, left) is coincident with the
position of the pulsar \psr. The emission roughly in the south is found to be coincident with the \FERMI\ source \twofgl\ and very
close to the radio source \pmn\ \citep[see][]{paper:radio_pmn_survey}; \twofgl\ is one of the three point-like sources in the 2FGL \FERMI\ catalog\footnote{The subsequent 3FGL \FERMI\ catalog \citep{Fermi3FGL} models \velajr\ as an extended source (\threefgl) and does not list a counterpart.} associated with \velajr\ as indicated in \citet{paper:fermi_ext_sources} and \citet{paper:fermi_2fgl_cat}.
This morphology is also seen in radio and \xr\ maps \citep{proc:vjr_radio_xrays} where most of the emission comes from the NW rim of the shell
with some local enhancements toward the south.
The azimuthal profile for the same region of the \ROSAT\ All Sky Survey for energies larger than 1.3 keV was derived as well (red points of Fig.~\ref{fig:az_prof}, right). The profiles of both instruments show a similar trend with a less pronounced enhancement toward \psr\ in the case of the \xrs. In addition, the \xr\ profile shows a clear local minimum at $\sim\!\!300\dg$, marking a break in the NW rim region of the shell. This discontinuity of the shell is not as pronounced in the \gr\ emission.
%

\subsection{Gamma-ray spectrum of the whole supernova remnant}\label{sec:spectrum}

The increased data set for \velajr\ enables a deeper study of the emission spectrum of the entire SNR.
After data quality selection, the available exposure only amounts to 21.0\,h since a sizable amount of the
available observations is not usable for spectrum determination because of the large extent of the \on\ region and
the exclusion region for Vela X. Many of the smaller \on\ regions used in the
spatially resolved spectroscopy analysis in Sect.~\ref{sec:spec_morph} achieve a better data efficiency.
The statistics and properties of the \on\ region encompassing the whole SNR are shown in row 2 of Table~\ref{tab:stats}.
The spectrum was calculated in the energy range from 0.3\,TeV to 30\,TeV with the forward-folding technique
assuming the three models listed in Table~\ref{tab:spectrum_models}: a plain power law, a curved power law, and a power law
with exponential cutoff. In the latter model, the parameter $\lambda = 1/E_\mathrm{cut}$ is used in the fit because its
error has a more Gaussian-like distribution than that of $E_\mathrm{cut}$.

\begin{table}[!htb]
  \centering
  \begin{tabular}{lll}
  \hline
  model & formula & parameters \\
  \hline
  \rule{0pt}{2ex}
  PL   & $ \mathrm{d}\Phi/\mathrm{d}E = \Phi_0 (E/E_0)^{-\Gamma} $                         & $ \Phi_0,\, \Gamma $                     \\
  CPL  & $ \mathrm{d}\Phi/\mathrm{d}E = \Phi_0 (E/E_0)^{-\Gamma - \beta \log(E/E_0)} $     & $ \Phi_0,\, \Gamma,\, \beta $            \\
  ECPL & $ \mathrm{d}\Phi/\mathrm{d}E = \Phi_0 (E/E_0)^{-\Gamma} \exp(-E/E_\mathrm{cut}) $ & $ \Phi_0,\, \Gamma,\, 1/E_\mathrm{cut} $ \\
  \hline
  \end{tabular}
  \caption[Spectrum models]{Spectrum models. For each model, the formula and the fit parameters are shown. The models are power law (PL), curved power law (CPL, also known as logarithmic parabola), and power law with exponential cutoff (ECPL).}
  \label{tab:spectrum_models}
\end{table}

The best parameters found for all three models are presented in Table~\ref{tab:spectrum_fits_curvature} (central section).
A likelihood ratio test \citep[based on the Wilks theorem from][]{paper:wilks_theorem} is used to select the model that describes the data best.
The null hypothesis is the power law model and the alternative hypothesis is either the curved power law or power law with exponential cutoff model.
Both alternative models are allowed to fall back into the power law during the fit, thereby fulfilling the nested hypotheses requirement for the test.

The results of the likelihood ratio tests are shown in the right section of Table~\ref{tab:spectrum_fits_curvature}. The
power law model is rejected at the $7.3\sigma$ level by the curved power law model and at the $7.7\sigma$ level by the power law with exponential cutoff model. This demonstrates that a curved
spectrum is clearly preferred over a plain power law, implying that an intrinsic curvature exists in the spectrum of \velajr.
Since the power law with exponential cutoff model shows the highest significance, it will be used in the following as the model describing the data best.
More complex models, such as power law with sub- or superexponential cutoff, were tested as well, but these models are
not significantly better than any of the two curved models with three parameters.

\begin{table*}[!htb]
  \centering
  \begin{tabular}{l|cccc|ccc}
  \hline
  model & $ \Phi_0\,[\UNITSwoS{cm^{-2} s^{-1} TeV^{-1}}]$ & $\Gamma$ & $\beta$ & $E_\mathrm{cut}\,[\UNITSwoS{TeV}]$ & $\log \lik$ & NFP & significance \\
  \hline
  PL   & $(27.4 \pm 0.9) \times 10^{-12}$ & $2.30 \pm 0.03$ & n/a            & n/a           & $-51.717$ & 2 & n/a        \\
  CPL  & $(28.8 \pm 1.1) \times 10^{-12}$ & $1.89 \pm 0.07$ & $0.23 \pm 0.04$ & n/a           & $-24.567$ & 3 & $7.3\sigma$ \\
  ECPL & $(32.2 \pm 1.5) \times 10^{-12}$ & $1.81 \pm 0.08$ & n/a            & $6.7 \pm 1.2$  & $-21.623$ & 3 & $7.7\sigma$ \\
  \hline
  \end{tabular}
  \caption[Fit parameters]{Parameters of the spectral fits of the entire SNR and results of the likelihood ratio tests for each of the models from Table~\ref{tab:spectrum_models}.
  The central section of the table lists the fit parameters.
  In the case of the power law with exponential cutoff, $E_\mathrm{cut}$ is shown instead of the fitted parameter $\lambda = 1/E_\mathrm{cut}$. The quoted errors represent $1\sigma$ statistical uncertainties. In all cases the reference energy $E_0$ was chosen to be 1\,TeV.
  The right section of the table shows the results of the likelihood ratio test in order to check for the existence of a curvature in the spectrum. The logarithm of the likelihood \lik, the number of free parameters NFP, and the equivalent significance of the probability of the test in Gaussian standard deviations $\sigma$ are shown for each fitted model. The test results apply to the comparison of a certain model with respect to the simpler (i.e., power law) model.}
  \label{tab:spectrum_fits_curvature}
\end{table*}

\begin{figure}[!htb]
  \centering
  \includegraphics[width=\hsize]{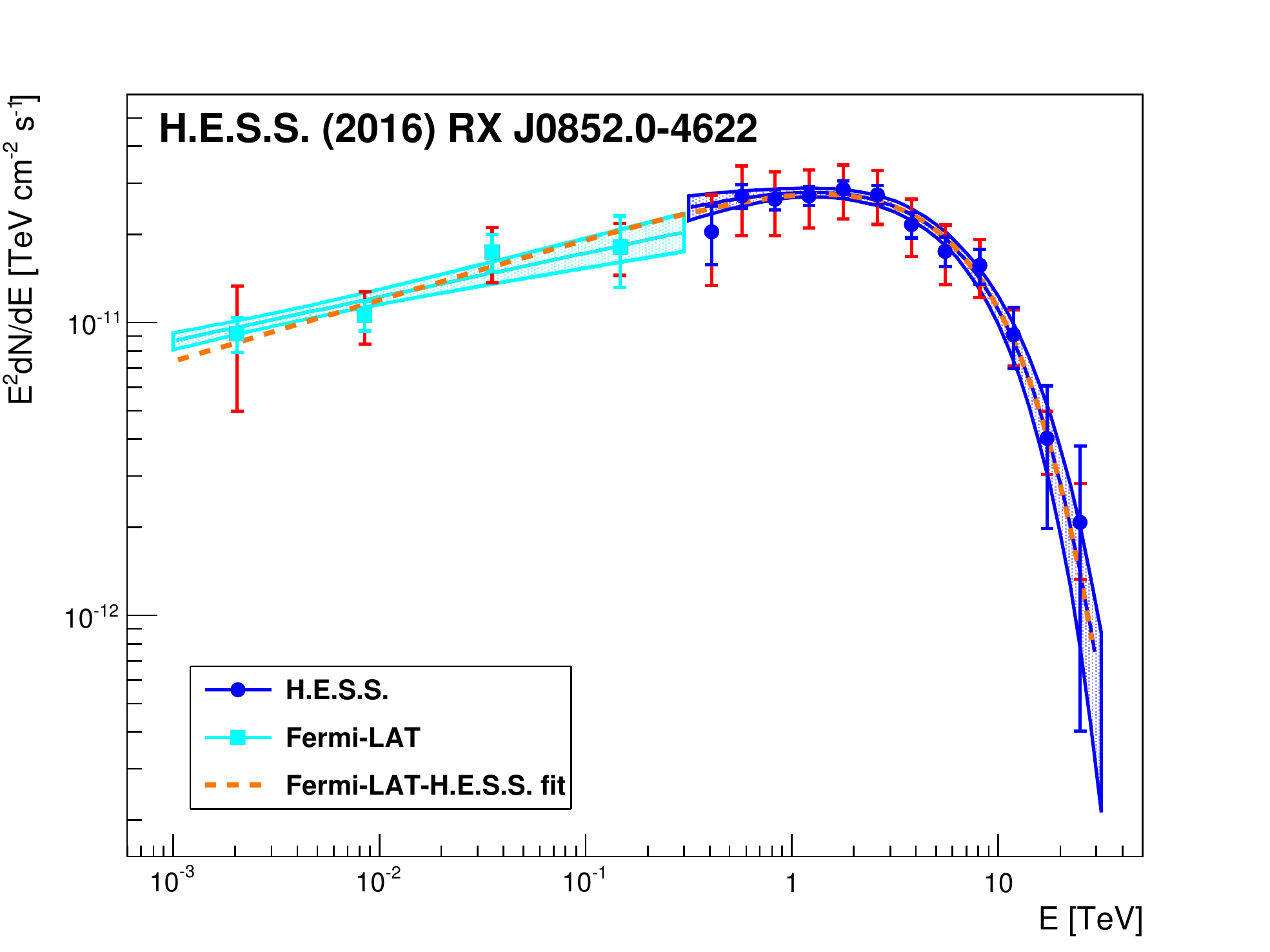}
  \caption[\HESS\ and \FERMI\ spectra for the entire SNR with statistical and systematic errors and simultaneous fit]{\HESS\ and \FERMI\ spectra for \velajr\ with statistical and systematic errors. The \FERMI\ measurement was taken from \citet{paper:vjr_fermi}. The figure shows the spectral points with $1\sigma$ statistical (light and dark blue lines for \FERMI\ and \HESS, respectively) and systematic (red lines in both cases) uncertainties together with the spectral fits and their $1\sigma$ statistical uncertainty (shaded bands). In addition, the simultaneous \FERMI-\HESS\ fit function is also shown (orange dashed line).}
  \label{fig:spectrum_hess_fermi}
\end{figure}

\begin{table*}[!htb]
  \centering
  \begin{tabular}{ll|l|l}
  \hline
  parameter & & \HESS & \FERMI-\HESS\ fit \\
  \hline
  \rule{0pt}{2ex}
  $\Phi_0$ & [$10^{-12} \UNITS{cm^{-2} s^{-1} TeV^{-1}}$] & $32.2 \pm 1.5\stat \pm 7.1\syst$ & $31.6 \pm 1.4\stat \pm 7.6\syst$ \\
  $\Gamma$    &      & $1.81 \pm 0.08\stat \pm 0.20\syst$ & $1.79 \pm 0.02\stat \pm 0.10\syst$ \\
  $E_\mathrm{cut}$ & [\UNITSwoS{TeV}]  & $6.7 \pm 1.2\stat \pm 1.2\syst$ & $6.6 \pm 0.7\stat \pm 1.3\syst$ \\
  $E_0$ & [\UNITSwoS{TeV}]  &  $1$ & $1$ \\
  $E_\mathrm{min}\!\!-\!\!E_\mathrm{max}$ & [\UNITSwoS{TeV}]  & $0.3\!\!-\!\!30$  & $0.001\!\!-\!\!30$  \\
  $F(>\!\!1 \UNITS{TeV}) $ & [$10^{-12} \UNITS{cm^{-2} s^{-1}}$]   & $23.4 \pm 0.7\stat \pm  4.9\syst$ & $23.2 \pm 0.7\stat \pm  5.6\syst$ \\
  $F(0.3\!\!-\!\!30\UNITS{TeV})$ & [$10^{-12} \UNITS{cm^{-2} s^{-1}}$] & $84.1 \pm 4.3\stat \pm 21.7\syst$ & $81.7 \pm 2.6\stat \pm 19.6\syst$ \\
  \hline
  \end{tabular}
  \caption[Spectrum parameters and integral fluxes]{Fit parameters for the \HESS\ spectrum (central column)
  and the simultaneous \FERMI-\HESS\ spectral fit (right column) of \velajr.
  The parameters refer to the power law with exponential cutoff model (ECPL in Table~\ref{tab:spectrum_models}).
  The parameter $E_\mathrm{cut}$ is shown instead of the
  fitted parameter $\lambda = 1/E_\mathrm{cut}$.
  The integral fluxes above 1\,TeV ($F(>\!\!1 \UNITS{TeV})$) and in the fitted range from 0.3\,TeV to 30\,TeV ($F(0.3\!\!-\!\!30\UNITS{TeV})$) are also shown. For the \HESS\ spectrum, these two fluxes represent $\sim\!\!113\%$ and $\sim\!\!64\%$, respectively, of the flux of the Crab nebula in the same energy ranges.
  The quoted errors represent $1\sigma$ statistical and systematic uncertainties.}
  \label{tab:spectrum_fit_pars_integral_fluxes}
\end{table*}

The \HESS\ spectrum of \velajr\ is shown in Fig.~\ref{fig:spectrum_hess_fermi} together with the \FERMI\ spectrum from \citet{paper:vjr_fermi}.
The \HESS\ spectral points and fit parameters are shown in Tables~\ref{tab:spectrum_points} and \ref{tab:spectrum_fit_pars_integral_fluxes} (central column), respectively.
The spectrum derived in this work represents a flux level that is 50\% higher than that derived in previous publications \citep{paper:vjr_hess_paper1,paper:vjr_hess_paper2}. A careful study has revealed that the lower flux found in the earlier analyses was due to a lack of correction for the degradation of the telescope reflectivities when estimating \gr\ energies and fluxes.
The revised flux makes \velajr\ the brightest steady source in the sky above 1\,TeV with $F(>\!\!1 \UNITS{TeV}) = (23.4 \pm 0.7\stat \pm 4.9\syst) \times 10^{-12} \UNITS{cm^{-2} s^{-1}}$
(a flux $\sim\!\!13\%$ larger than the flux of the Crab nebula\footnote{Crab nebula fluxes in this work are calculated using the spectrum from \citet{paper:crab_meyer}.} in the same energy range)
and results in a smooth connection of the GeV (\FERMI) and TeV (\HESS) spectra (cf.~Fig.~\ref{fig:spectrum_hess_fermi}).
Indeed, the spectral break that is visible at the $5.4\sigma$ level (statistical; or $1.6\sigma$ systematic) between the
GeV ($\Gamma = 1.85 \pm 0.06\stat\ ^{+ 0.18\syst} _{- 0.19\syst}$) and the TeV ($\Gamma = 2.24 \pm 0.04\stat \pm 0.15\syst$) spectra
in \citet{paper:vjr_fermi} with the current measurement from Table~\ref{tab:spectrum_fit_pars_integral_fluxes} (central column) is only $0.4\sigma$ (statistical), i.e.,~nonexistent.

Using both the \FERMI\ and \HESS\ measurements, the ambiguity between the power law with exponential cutoff and curved power law models can be solved.
Indeed, a simultaneous fit of the \FERMI\ and \HESS\ data points yields a $\chi^2$ fit probability of 94\% for a power law with exponential cutoff model,
whereas the probability for the curved power law is $3.1 \times 10^{-7}$. The simultaneous \FERMI-\HESS\ fit using the power law with exponential cutoff model is shown in Fig.~\ref{fig:spectrum_hess_fermi} with an orange dashed line and the fit parameters are shown in the rightmost
column of Table~\ref{tab:spectrum_fit_pars_integral_fluxes}.
We determined the systematic uncertainties of the parameters of the simultaneous fit
from the variations observed when moving the \FERMI\ points down (up) and the \HESS\ points up (down) by
one standard deviation of their respective systematic uncertainties, which tests the systematic uncertainty of the spectral index and cutoff energy.
We also moved all data points down or up simultaneously to test the systematic uncertainty of the normalization.

A comparison of the parameter values of the simultaneous \FERMI-\HESS\ fit and of
the \HESS\ only spectrum (cf.~Table~\ref{tab:spectrum_fit_pars_integral_fluxes}) shows good agreement.
The simultaneous fit shows smaller statistical uncertainties for all parameters (especially for the spectral index and cutoff energy), a smaller systematic uncertainty for the spectral index,
a larger systematic uncertainty in the normalization, and a similar systematic uncertainty in the cutoff energy.
Thus, the simultaneous fit is able to better determine the spectral index and cutoff energy at the cost of a higher systematic uncertainty in the flux.
The latter is due to the larger systematic error in the \FERMI\ flux measurement ($\sim\!\!30\!\!-\!\!35\%$).
The simultaneous fit also shows a smaller uncertainty for the low-energy part of the spectrum ($E < 1 \UNITS{TeV}$), as shown by the smaller statistical uncertainty in the $F(0.3\!\!-\!\!30\UNITS{TeV})$ quantity, in contrast to the same uncertainty for the $F(>\!\!1 \UNITS{TeV})$ quantity.
%

\subsection{Spatially resolved spectroscopy}\label{sec:spec_morph}

\begin{figure*}[!htb]
  \centerline{
    \includegraphics[width=2.4in]{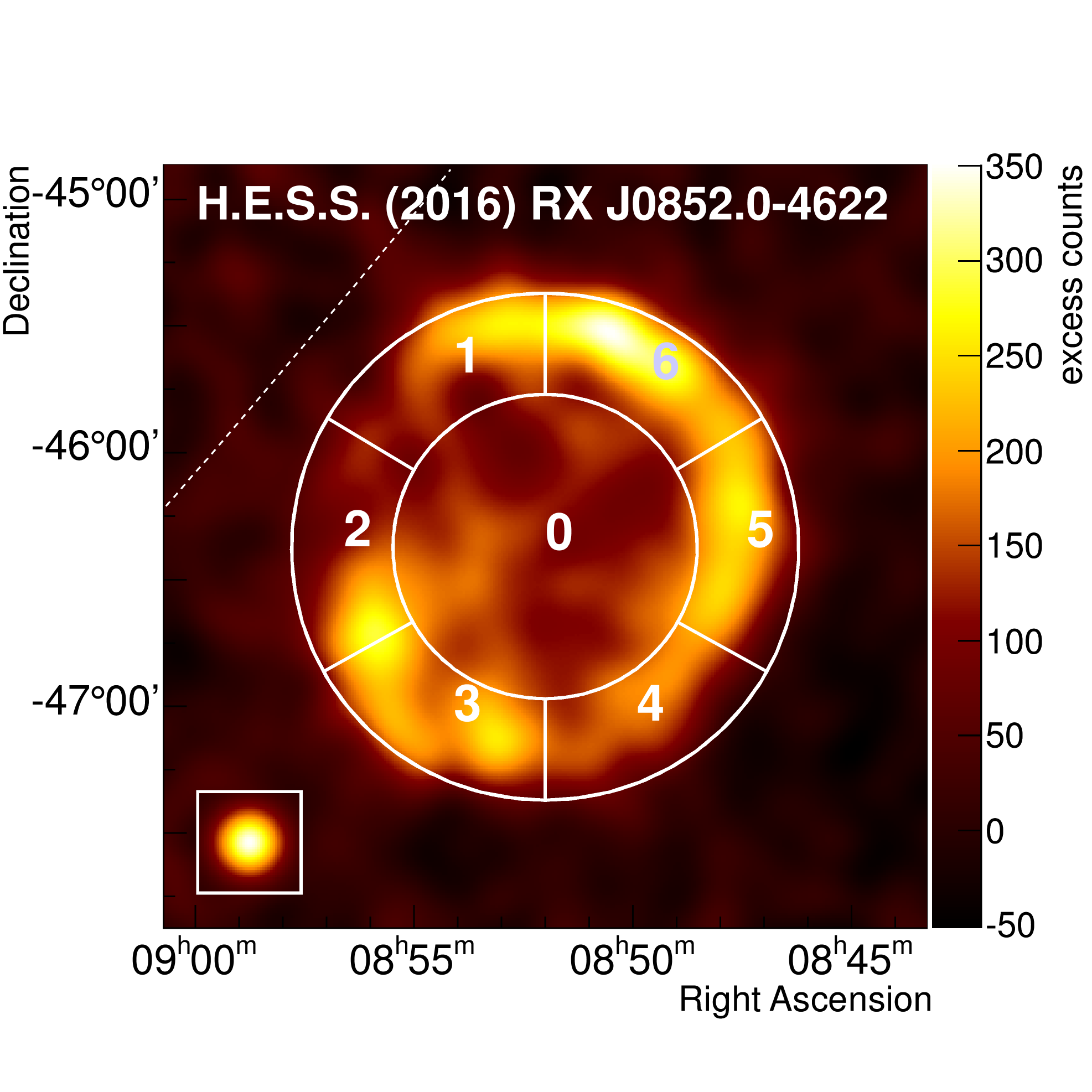}
    \hfil
    \includegraphics[width=2.4in]{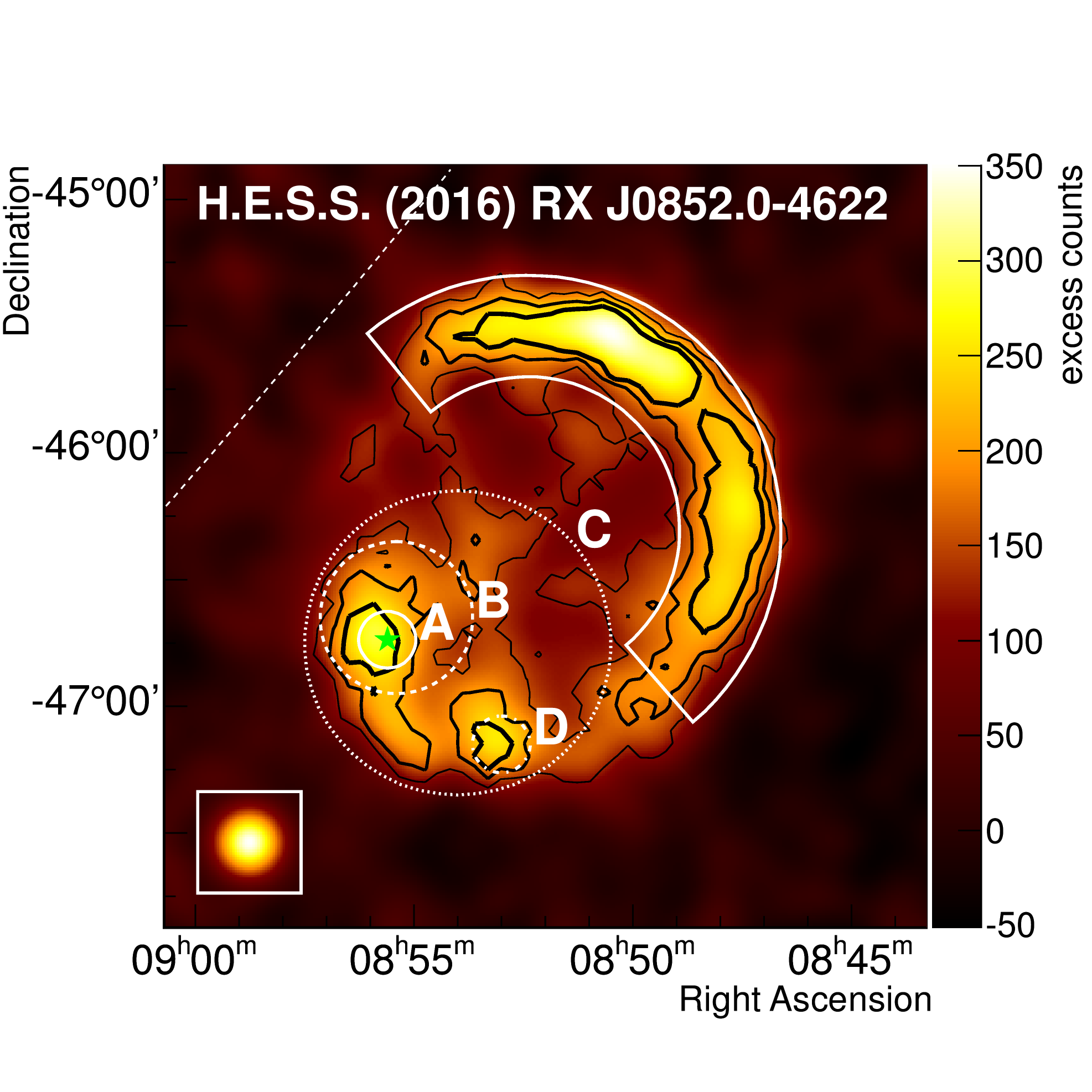}
    \hfil
    \includegraphics[width=2.4in]{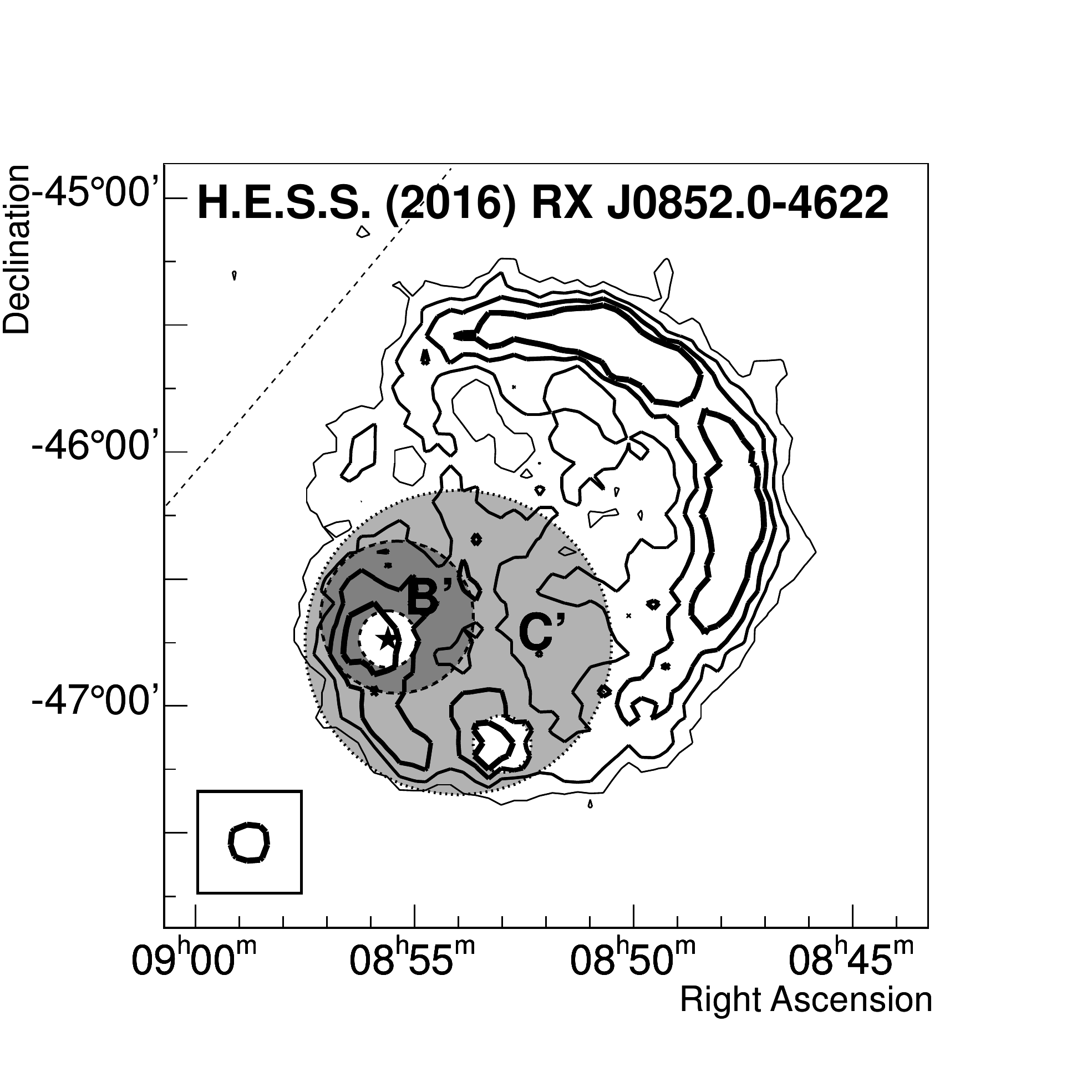}
  }
  \centerline{
    \includegraphics[width=2.4in]{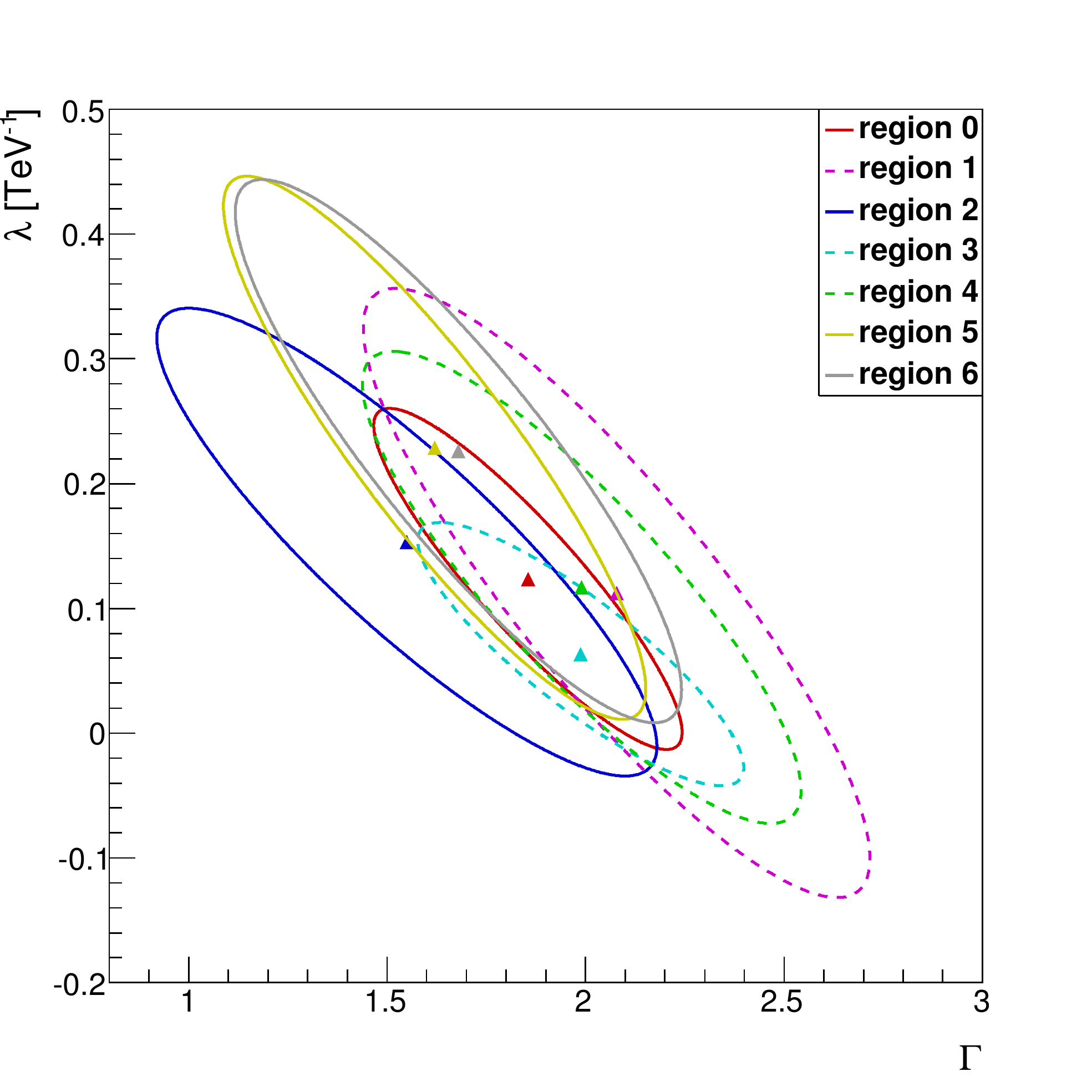}
    \hfil
    \includegraphics[width=2.4in]{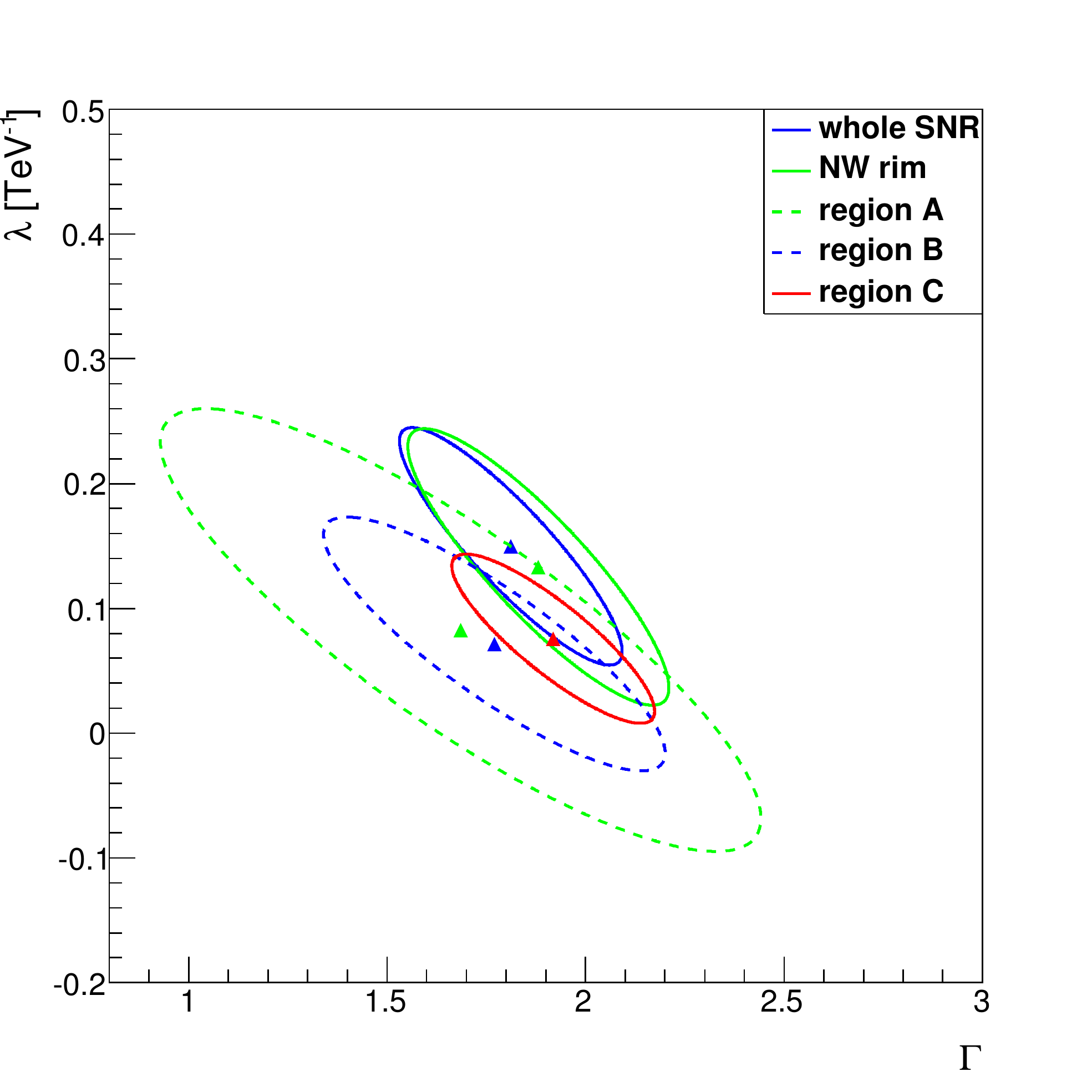}
    \hfil
    \includegraphics[width=2.4in]{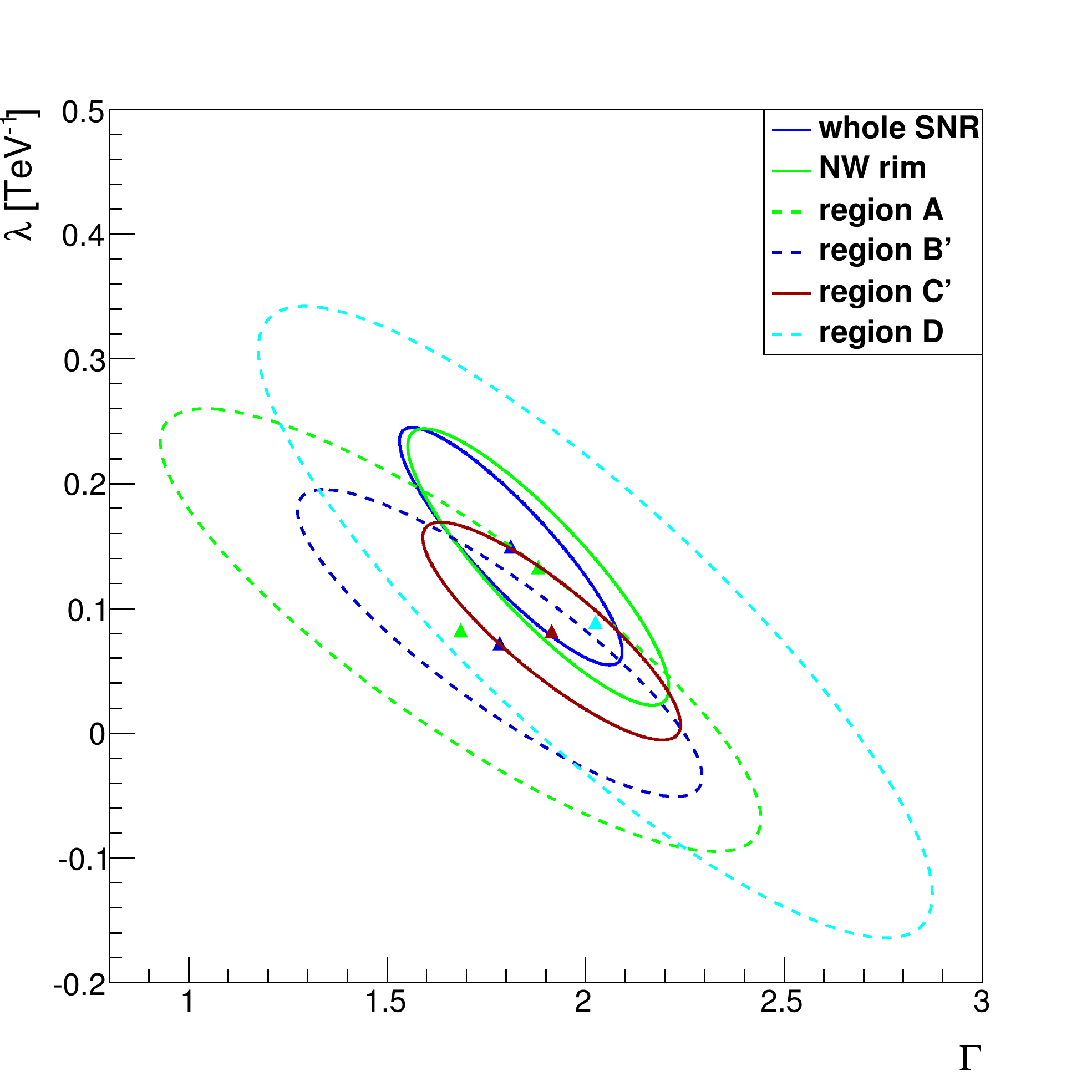}
  }
  \caption[Spatially-resolved spectroscopy maps and spectral contours]{Top figures represent skymaps (smoothed exposure-corrected excess map from Fig.~\ref{fig:skymaps} for the top left and center pads, significance contour map for the right pad) together with various regions used for the study of the spatially resolved spectroscopy of the \velajr\ region.
  Top left: regions 0 for the central part of the SNR and 1--6 for the shell are shown in white.
  Center: regions A, B, and C around \psr, region D around the southern enhancement in the azimuthal profile, and the region used for the NW rim are shown in white. The \HESS\ significance contours at 5, 7, and $9\sigma$ are shown in black with increasing line width for increasing significance. The position of \psr\ is denoted by the green star.
  Top right: regions B$'$ (dark gray shaded area) and C$'$ (light gray shaded area) around \psr\ are shown. The \HESS\ significance contours at 3, 5, 7, and $9\sigma$ are shown in black with increasing line width for increasing significance. The position of \psr\ is denoted by the black star.
  Bottom figures represent the error contour plots of $\lambda = 1/E_\mathrm{cut}$ vs. $\Gamma$ for the spectra of the regions in the excess maps assuming a power law with exponential cutoff model. For each region, the fitted value is indicated with a triangle, while the ellipse indicates the contour of the $3\sigma$ confidence level statistical uncertainty. Dashed lines are used for the regions where the cutoff significance is below $3\sigma$.
  Bottom left: spectra of the regions 0 to 6 are represented.
  Bottom center: spectra of the regions A, B, and C are represented. In addition, the contours of the analysis of the whole SNR and NW rim are shown. Since the spectra of regions A--C are correlated because the regions overlap, the corresponding contours are not meant to be compared to each other, but to the contours of the NW rim.
  Bottom right: spectra of the regions A, B$'$, C$'$, and D are represented. In addition, the contours of the analysis of the whole SNR and the NW rim are shown.}
  \label{fig:spec_morph_maps_contours}
\end{figure*}

The increased data set on \velajr\ and the size of the SNR allow for the first time a
spatially resolved spectroscopy, i.e.,~the derivation of spectra for subregions. For this purpose,
we defined different subregions of \velajr\ following its \gr\ morphology and the location
of the PWN around \psr\ and the southern enhancement. As shown in Fig.~\ref{fig:spec_morph_maps_contours} (top left), we divided the SNR into
a central part (region 0) and six regions (1--6) in the annulus covering the shell to test for spectral variations across the SNR. An additional
region encompasses the $5\sigma$ significance contour around the bright NW rim (cf.~Fig.~\ref{fig:spec_morph_maps_contours}, top center).
Three circular regions around the position of \psr\ (regions A, B, and C in Fig.~\ref{fig:spec_morph_maps_contours}, top center)
are used to estimate a possible PWN flux. Region A supposes a point-like source centered at the pulsar position.
Region B encompasses the $7\sigma$ contour around the pulsar excluding the elongation to the south.
Region C roughly encompasses the $5\sigma$ contour around the pulsar.
Region D (cf.~Fig.~\ref{fig:spec_morph_maps_contours}, top center) is adapted to a hypothetical point-like source for the southern enhancement observed in the
skymap and the azimuthal profile (cf.~Fig.~\ref{fig:az_prof}).
In addition, the regions
$\mbox{B}' = \mbox{B}\setminus\mbox{A}$ (i.e.,~region B, excluding region A) and
$\mbox{C}' = \mbox{C}\setminus\mbox{B}\setminus\mbox{D}$ (i.e.,~region C, excluding regions B and D)
are used to search for a softening of the spectrum of the possible TeV PWN around \psr\ (shaded areas in Fig.~\ref{fig:spec_morph_maps_contours}, top right).
The specific parameters for all regions used are defined in Table~\ref{tab:reg_def}.

Spectral analyses similar to those presented in Sect.~\ref{sec:spectrum} were performed for each of the regions.
The event statistics of all analyses are presented in Table~\ref{tab:stats}. All analyzed regions show a clear signal with
significances between $7.7\sigma$ and $29.0\sigma$. The spectral parameters of a power law with exponential cutoff
fitted to the data of each analyzed region in the energy range from 0.3\,TeV to 30\,TeV
are shown in Table~\ref{tab:spec_morph_spectral_pars}. The significance of
the cutoff is listed in the rightmost column.

\begin{table*}[!htb]
  \centering
  \addtolength{\tabcolsep}{-1pt}
  \begin{tabular}{lccccccc}
  \hline
  region    & $\Phi_0$                                     & $\Gamma$                   & $E_\mathrm{cut}$         & $F(>\!\!1 \UNITS{TeV})$             & $F(0.3\!\!-\!\!30 \UNITS{TeV})$     & sign        \\
            & [$10^{-12} \UNITS{cm^{-2} s^{-1} TeV^{-1}}$] &                            & [$\UNITSwoS{TeV}$]       & [$10^{-12} \UNITS{cm^{-2} s^{-1}}$] & [$10^{-12} \UNITS{cm^{-2} s^{-1}}$] &             \\
  \hline
  \hline
  whole SNR & $32.2  \pm 1.5  (\pm 7.1)$                   & $1.81 \pm 0.08 (\pm 0.20)$ & $ 6.7 \pm 1.2 (\pm 1.2)$ & $23.4  \pm 0.7  (\pm 4.9 )$         & $84.1 \pm 4.3 (\pm 21.7)$           & $7.7\sigma$ \\
  \hline
  NW rim    & $12.4  \pm 0.7  (\pm 3.1)$                   & $1.88 \pm 0.10 (\pm 0.20)$ & $ 7.5 \pm 1.8 (\pm 1.5)$ & $ 8.9  \pm 0.4  (\pm 2.2 )$         & $33.7 \pm 2.0 (\pm 8.4 )$           & $5.6\sigma$ \\
  \hline
  0         & $ 8.9  \pm 0.6  (\pm 2.2)$                   & $1.85 \pm 0.11 (\pm 0.20)$ & $ 8.1 \pm 2.6 (\pm 1.6)$ & $ 6.7  \pm 0.3  (\pm 1.7 )$         & $24.1 \pm 1.6 (\pm 6.0 )$           & $4.5\sigma$ \\
  1         & $ 3.5  \pm 0.4  (\pm 0.9)$                   & $2.08 \pm 0.19 (\pm 0.20)$ & $ 8.9 \pm 5.6 (\pm 1.8)$ & $ 2.34 \pm 0.20 (\pm 0.59)$         & $10.5 \pm 1.1 (\pm 2.6 )$           & $2.1\sigma$ \\
  2         & $ 2.4  \pm 0.3  (\pm 0.6)$                   & $1.55 \pm 0.18 (\pm 0.20)$ & $ 6.5 \pm 2.3 (\pm 1.3)$ & $ 2.12 \pm 0.16 (\pm 0.53)$         & $ 5.9 \pm 0.7 (\pm 1.5 )$           & $3.9\sigma$ \\
  3         & $ 4.0  \pm 0.3  (\pm 1.0)$                   & $1.99 \pm 0.12 (\pm 0.20)$ & $15.8 \pm 7.7 (\pm 3.2)$ & $ 3.19 \pm 0.19 (\pm 0.80)$         & $12.1 \pm 1.0 (\pm 3.0 )$           & $2.3\sigma$ \\
  4         & $ 3.8  \pm 0.4  (\pm 1.0)$                   & $1.99 \pm 0.16 (\pm 0.20)$ & $ 8.6 \pm 4.0 (\pm 1.7)$ & $ 2.64 \pm 0.20 (\pm 0.66)$         & $10.9 \pm 1.1 (\pm 2.7 )$           & $2.7\sigma$ \\
  5         & $ 4.4  \pm 0.4  (\pm 1.1)$                   & $1.62 \pm 0.15 (\pm 0.20)$ & $ 4.4 \pm 1.2 (\pm 0.9)$ & $ 2.99 \pm 0.17 (\pm 0.75)$         & $ 9.9 \pm 0.8 (\pm 2.5 )$           & $5.5\sigma$ \\
  6         & $ 5.2  \pm 0.5  (\pm 1.3)$                   & $1.68 \pm 0.16 (\pm 0.20)$ & $ 4.4 \pm 1.2 (\pm 0.9)$ & $ 3.43 \pm 0.22 (\pm 0.86)$         & $12.0 \pm 1.1 (\pm 3.0 )$           & $4.9\sigma$ \\
  \hline
  A         & $ 0.64 \pm 0.10 (\pm 0.16)$                  & $1.69 \pm 0.22 (\pm 0.20)$ & $12.1 \pm 7.6 (\pm 2.4)$ & $ 0.62 \pm 0.07 (\pm 0.15)$         & $ 1.8 \pm 0.3 (\pm 0.4 )$           & $1.8\sigma$ \\
  B         & $ 2.34 \pm 0.19 (\pm 0.59)$                  & $1.77 \pm 0.13 (\pm 0.20)$ & $14.0 \pm 5.8 (\pm 2.8)$ & $ 2.18 \pm 0.14 (\pm 0.55)$         & $ 6.6 \pm 0.7 (\pm 1.7 )$           & $2.7\sigma$ \\
  C         & $10.0  \pm 0.5  (\pm 2.5 )$                  & $1.92 \pm 0.07 (\pm 0.20)$ & $13.2 \pm 3.4 (\pm 2.6)$ & $ 8.2  \pm 0.3  (\pm 2.1 )$         & $29.5 \pm 1.6 (\pm 7.4 )$           & $4.5\sigma$ \\
  D         & $ 0.81 \pm 0.12 (\pm 0.20)$                  & $2.02 \pm 0.25 (\pm 0.20)$ & $11.2 \pm 9.3 (\pm 2.2)$ & $ 0.59 \pm 0.07 (\pm 0.15)$         & $ 2.4 \pm 0.4 (\pm 0.6 )$           & $1.4\sigma$ \\
  B$'$      & $ 1.88 \pm 0.18 (\pm 0.47)$                  & $1.78 \pm 0.15 (\pm 0.20)$ & $13.9 \pm 6.9 (\pm 2.8)$ & $ 1.73 \pm 0.13 (\pm 0.43)$         & $ 5.4 \pm 0.6 (\pm 1.3 )$           & $2.3\sigma$ \\
  C$'$      & $ 6.7  \pm 0.4  (\pm 1.7 )$                  & $1.91 \pm 0.09 (\pm 0.20)$ & $12.2 \pm 3.8 (\pm 2.4)$ & $ 5.4  \pm 0.3  (\pm 1.4 )$         & $19.6 \pm 1.4 (\pm 4.9 )$           & $3.8\sigma$ \\
  \hline
  \end{tabular}
  \addtolength{\tabcolsep}{1pt}
  \caption[Spatially-resolved spectroscopy spectral parameters]{Spectral parameters for the spatially resolved spectroscopy of \velajr\ assuming a power law with exponential cutoff model (ECPL in Table~\ref{tab:spectrum_models}).
  The regions are defined in Table~\ref{tab:reg_def} and illustrated in Figs.~\ref{fig:skymaps}~and~\ref{fig:spec_morph_maps_contours} for the whole SNR and the rest of the regions, respectively.
  The parameter set of each region was derived for the corresponding spectral analysis listed in Table~\ref{tab:stats}.
  All parameter sets were derived for spectra in the energy range from 0.3\,TeV to 30\,TeV and assuming a reference energy for the fit $E_0$ of 1\,TeV; $F(>\!\!1 \UNITS{TeV})$ and $F(0.3\!\!-\!\!30\UNITS{TeV})$ represent the integral fluxes above 1\,TeV and in the fitted range from 0.3\,TeV to 30\,TeV, respectively. The last column lists the equivalent significance in Gaussian standard deviations $\sigma$ of the preference of a power law with exponential cutoff model with respect to a simple power law model, according to the likelihood ratio test.
  The quoted errors represent $1\sigma$ statistical (systematic in parentheses) uncertainties.}
  \label{tab:spec_morph_spectral_pars}
\end{table*}

The central part of the SNR (region 0) and the shell toward the bright NW rim show a preference for a cutoff in the spectrum (test significance greater than $3\sigma$); for the enhancement
toward \psr\ the significance of the cutoff is only 1.8$\sigma$ for the point-like region (region A) but it grows with the size of the
integration region (regions B and C).
As for spectral variations across the \velajr\ region, the shape of the spectrum of the NW rim is compatible with the spectrum of the whole SNR.
This spectrum is also compatible with the spectra of the rest of the SNR at the 2$\sigma$ level when comparing to either regions C or 0, which largely cover the rest of the SNR emission.
The spectra of the regions 0 to 6 do not show a clear deviation from the spectrum of the whole SNR ($\Gamma \sim 1.8$, $\lambda \sim 0.15 \UNITS{TeV^{-1}}$), as shown by the overlapping error contours in Fig.~\ref{fig:spec_morph_maps_contours} (bottom left); the variations of the spectral index $\Gamma$ are smaller than $\pm 2\sigma$.
Although region 2 seems to deviate from regions 1, 4, and 6, the effect is not significant, and indeed the weighted average of ($\Gamma$, $\lambda$) for regions 0 to 6 still yields a $\chi^2$ probability of 4.8\% (residual significance of $2.0\sigma$), which shows that all regions are compatible with a single ($\Gamma$, $\lambda$) pair.
The profile-likelihood method was applied to quantify the hypothesis that any of the seven regions 0 to 6
has a spectral index that differs by more than $\pm\Delta\Gamma$ from the spectral index calculated from
the remaining six regions. Depending on the region, the upper limits on $\Delta\Gamma$ calculated at the 95\% confidence level
are in the range of 0.25 to 0.40, which shows that the available statistics are not sufficient for
a sensitive search for spectral variations.
Similar values are found when using combinations of the regions from the top center and top right of Fig.~\ref{fig:spec_morph_maps_contours}.
Also, spectral variations on angular scales comparable to or below the angular
resolution of \HESS\ are not excluded.
The spectra of the regions A and C also do not show a deviation from the spectra of either the whole SNR or the NW rim, as shown by the contour plot of Fig.~\ref{fig:spec_morph_maps_contours} (bottom center). Region B shows some evidence of a harder spectrum with a larger cutoff energy compared to the NW rim (or the whole SNR); the significance for the cutoff in region B is not very high (only $2.7\sigma$) and the deviation of index and cutoff energy is
at the level of $3.5\sigma$ pretrials ($2.6\sigma$ post-trials\footnote{The post-trial significance takes into account that $\sim\!\!20$ regions were tested
in the search for spectral variations.}).
Figure~\ref{fig:spec_morph_maps_contours} (bottom right) shows that the spectral index of the possible TeV PWN remains basically constant when moving
farther away from \psr\ (region A to B$'$ to C$'$): no significant change in the spectral parameters is observed across the mentioned regions.
This plot also shows that the spectrum of region D is compatible with that of the NW rim. In addition, the spectrum of region B$'$, although it comprises a smaller exposure than that of region B, still shows marginal evidence for a difference with respect to the spectrum of the NW rim.
The significance of the effect is $3.1\sigma$ pretrials ($2.1\sigma$ post-trials).
Nevertheless, the weighted average of ($\Gamma$, $\lambda$) for regions A, B$'$, C$'$, D, and the NW rim still yields a $\chi^2$ probability of 6.6\% (residual significance of $1.8\sigma$).
%

\section{Discussion}\label{sec:interpretation}

In the following, we consider the implications of the new \HESS\ results on the understanding of the very high-energy \gr\ emission
from \velajr. The multiwavelength (MWL) data used in this section consist of
\begin{itemize}
  \item \PARKES\ radio data points from \citet{paper:vjr_radio_points};
  \item \ASCA\ \xr\ spectral fit from \citet{paper:vjr_hess_paper2};
  \item \FERMI\ GeV \gr\ points from \citet{paper:vjr_fermi};
  \item \HESS\ TeV \gr\ points from this work (cf.~Fig.~\ref{fig:spectrum_hess_fermi} and Table~\ref{tab:spectrum_points}).
\end{itemize}
The \HESS\ spectrum is not corrected for possible emission from a putative
TeV PWN around \psr\ (discussed in Sect.~\ref{interpretation:pwn}) since the expected flux
contribution is smaller than the systematic error on the flux normalization.
%

\subsection{Spectral variations}

The spatially resolved spectroscopy analysis of \velajr\ described in Sect.~\ref{sec:spec_morph} does not show a significant variation
of the spectral shape across the remnant. The regions that contain the PWN (cf.~Fig.~\ref{fig:spec_morph_maps_contours}; discussed in Sect.~\ref{interpretation:pwn})
show some evidence for spectral variations but the effect is still below $3\sigma$.
This lack of apparent spectral variations suggests that the parent particle population is essentially the same across the remnant; this in turn points
to similar properties of the SNR shock and hence to similar properties of the medium in which the shock was formed and is expanding.
The invariability of the \gr\ spectral
shape across the remnant enables a simple treatment of the MWL emission from \velajr,
using one particle population for the entire SNR emission.
This modeling obtains the average
properties of the SNR and its surrounding medium.
%

\subsection{Parent particle population}

\begin{figure*}[!htb]
  \centerline{
    \includegraphics[width=3.5in]{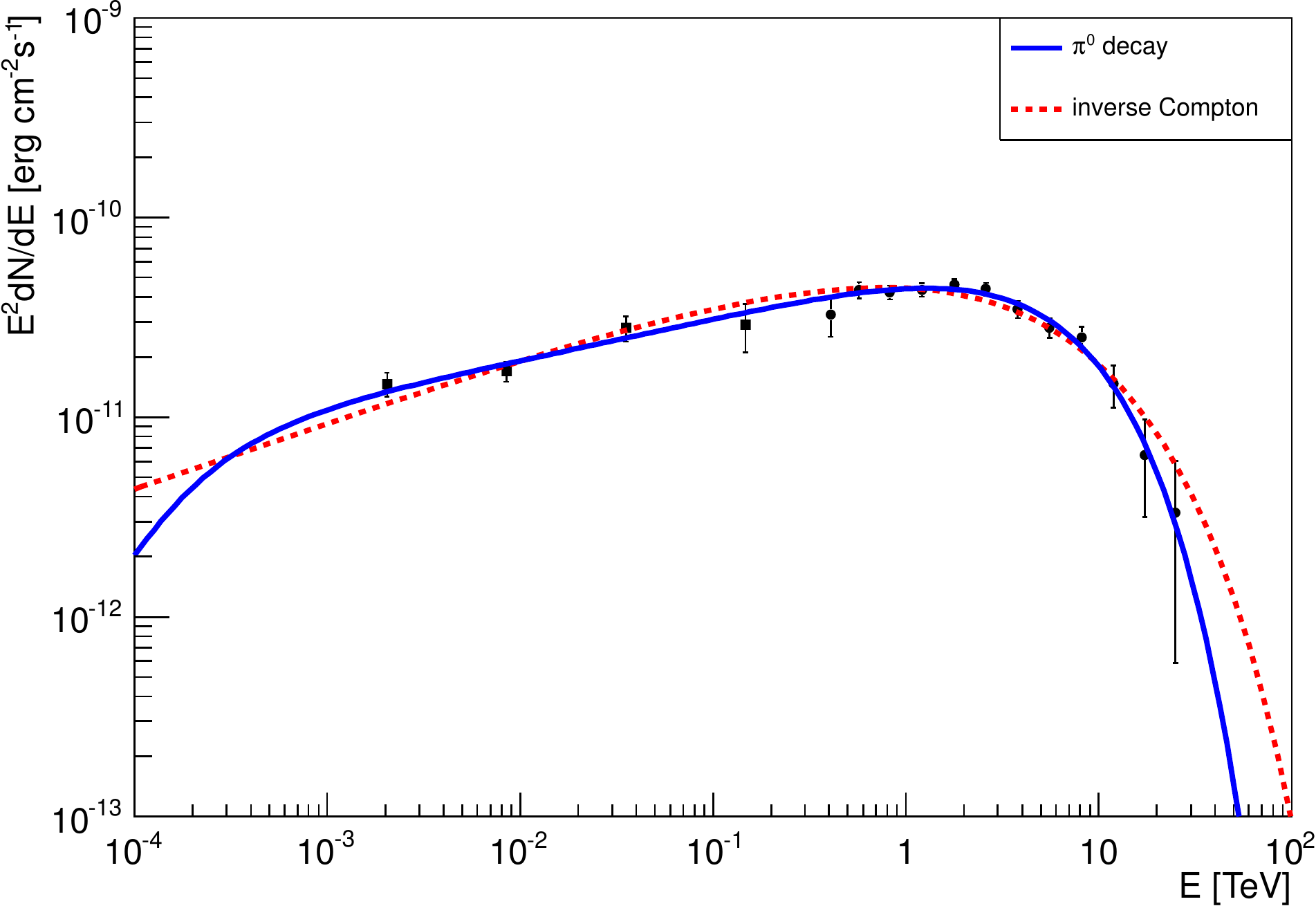}
    \hfil
    \includegraphics[width=3.5in]{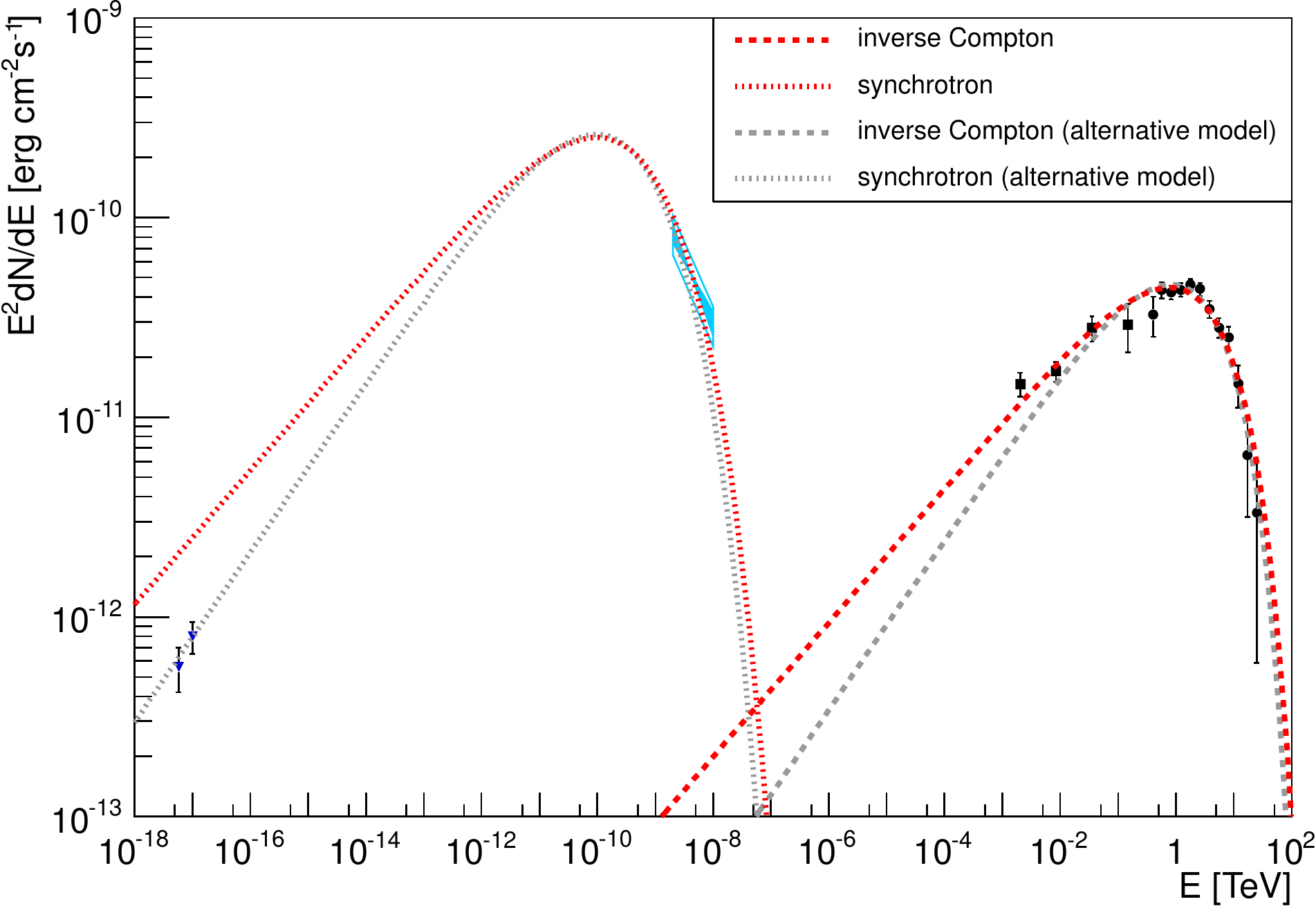}
  }
  \caption[SED models]{Left: spectral energy distribution of SNR \velajr\ in the GeV-TeV band. Filled squares reflect the emission detected by \FERMI,
while filled circles show the \HESS\ data. Error bars reflect statistical errors. The lines represent
fits of the leptonic (dashed red) and hadronic (solid blue) \gr\ emission models to the data.
Right: broadband spectral energy distribution of SNR \velajr\ in the leptonic scenario. The MWL data are indicated as follows: radio data are denoted with filled blue triangles,
\xr\ are indicated with cyan bowties, GeV \grs\ are indicated with filled squares, TeV \grs\ are denoted with filled circles. The error bars of the \FERMI\ and \HESS\ data points reflect
statistical errors only.
The filled blue bowtie represents the 90\% confidence level statistical uncertainty band on the \xr\ data; the open bowtie represents systematic uncertainties.
Red lines represent the emission from the electron population as obtained from the
fit of the GeV-TeV data (cf.~left figure) and gray lines represent the emission from a modified electron population where
all parameters were decreased by $0.6\sigma$ of the quadratically added statistical and systematic uncertainties. Dashed lines correspond to IC emission and dotted lines
correspond to synchrotron radiation.}
  \label{fig:gamma_sed}
\end{figure*}

The increased exposure and careful study of the systematic effects strongly improved the quality of the spectrum of the whole SNR, thereby resulting
in a smooth connection to the spectrum in the GeV band and in the determination of a clear cutoff (cf.~Sect.~\ref{sec:spectrum}).
The smooth connection to the \FERMI\ spectrum enables the
study of the combined GeV-TeV spectrum, which in turn provides an opportunity to extract directly from the observational data the present-time
parent particle population that is responsible for the \gr\ emission in leptonic and hadronic scenarios.
The advantage of this approach
is that the spectral shape of the present-time parent particle population can be obtained without assumptions made on the SNR evolution,
its hydrodynamics, properties of the local magnetic field, and energy losses that accelerated particles undergo.
In this procedure we fit the \FERMI\ and \HESS\ data points and their respective statistical errors
(cf.~Fig.~\ref{fig:gamma_sed}, left) with the emission from parent electron or proton populations following a power law with an exponential cutoff
\begin{equation}\label{eq:parent_pop_spectrum}
  N_{\mathrm{p},\mathrm{e}}(E) = \frac{N_{0,\,\mathrm{p},\mathrm{e}}}{4 \pi d^2} \left(\frac{E}{1\,\mathrm{TeV}}\right)^{-p_{\mathrm{p},\mathrm{e}}} \exp \left({-\frac{E}{E_{\mathrm{cut},\,\mathrm{p},\mathrm{e}}}}\right),
\end{equation}
where $N_{0,\,\mathrm{p},\mathrm{e}}$ is the normalization at $1$\,TeV, $p_{\mathrm{p},\mathrm{e}}$ is the spectral index, $E_{\mathrm{cut},\,\mathrm{p},\mathrm{e}}$ is the cutoff energy, and $d$ is the distance to the SNR. The subscripts $\mathrm{p}$ (for protons) and $\mathrm{e}$ (for electrons) denote the hadronic and leptonic scenarios, respectively.
In order to evaluate the systematic uncertainty of our results, we also fit the model when we systematically shifted the \FERMI\ and \HESS\ points in an analogous way to the simultaneous \FERMI-\HESS\ spectral fit in Sect.~\ref{sec:spectrum}.
The distance of 750\,pc is adopted in these calculations. Uncertainties
on the distance estimate only impact flux normalization, and hence the estimate of the total energy in particles, but these uncertainties do not influence the
spectral shape, i.e.,~the spectral index and cutoff energy. The spectral index is determined by the smooth connection of the \FERMI\ and \HESS\
spectra, while the cutoff energy of the accelerated particles is constrained by the cutoff in the \HESS\ \gr\ spectrum. The derivation of the present-time
parent particle population is very important for further modeling of the source.
Hadronic and leptonic scenarios can be tested based on the attainability of the present-time particle population considering the physical properties of the SNR and its ambient medium.

\begin{table}[!htb]
  \centering
  \addtolength{\tabcolsep}{-2pt}
  \begin{tabular}{llll}
  \hline
  scenario & parameter & & value
  \\
  \hline
  \hline
  		  	& $\frac{N_{0,\,\mathrm{e}}}{4 \pi d^2}$
  		  	& [$10^2\,\mathrm{TeV}^{-1}\mathrm{cm}^{-2}$]
  			& $7.8 \pm 0.6 (\pm 3.1)$
  \\
  leptonic 	        & $p_\mathrm{e}$
  			&
  			& $2.33 \pm 0.03 (\pm 0.33)$
  \\
  		 	& $E_{\mathrm{cut},\,\mathrm{e}}$
  		 	& [TeV]
  			& $27 \pm 1 (\pm 12)$
  \\
  		 	& $W_\mathrm{e}$
  		 	& [$10^{47} \mathrm{erg}$]
  			& $4.1 \pm 0.3 (\pm 1.7)$
  \\
  \hline
  		 	& $\frac{N_{0,\,\mathrm{p}}}{4 \pi d^2} \left[n\right]^{-1}$
  		 	& [$10^4\,\mathrm{TeV}^{-1}\mathrm{cm}^{-2}$]
  			& $7.8 \pm 0.3 (\pm 2.0)$
  \\
  hadronic 	        & $p_\mathrm{p}$
  			&
  			& $1.83 \pm 0.02 (\pm 0.11)$
  \\
  		 	& $E_{\mathrm{cut},\,\mathrm{p}}$
  		 	& [TeV]
  			& $55 \pm 6 (\pm 13)$
  \\
  		 	& $W_\mathrm{p} \left[n\right]^{-1}$
  		 	& [$10^{49} \mathrm{erg}$]
  			& $7.1 \pm 0.3 (\pm 1.9)$
  \\
  \hline
  \end{tabular}
  \addtolength{\tabcolsep}{2pt}
  \caption{Best parameter sets of the fit of the parent particle distribution of Eq.~(\ref{eq:parent_pop_spectrum}) in the leptonic (subscript $\mathrm{e}$ for electrons)
  and hadronic (subscript $\mathrm{p}$ for protons) scenarios using solely the \gr\ data. $N_{0,\,\mathrm{e},\mathrm{p}}$ represents the differential particle number at 1\,TeV,
  $p_{\mathrm{e},\mathrm{p}}$ the spectral index, $E_{\mathrm{cut},\,\mathrm{e},\mathrm{p}}$ the cutoff energy, $d$ the distance to the
  SNR (using 750 pc), $n$ the density of the ambient medium (in cm$^{-3}$), and $W_{\mathrm{e},\mathrm{p}}$ the total energy in accelerated particles above $100 \UNITS{GeV}$ for electrons in the leptonic model
  and above $1 \UNITS{GeV}$ for protons in the hadronic model. The quoted errors represent $1\sigma$ statistical (systematic in parentheses) uncertainties.
  }
  \label{tab:ParamsParentParticles}
\end{table}
%

\subsubsection{Leptonic scenario}

In the leptonic scenario, we assume that the GeV-TeV \gr\ emission from \velajr\ is dominated by the IC emission from relativistic
electrons scattered on ambient radiation fields. Besides the cosmic microwave background (CMB), local infrared (IR) and optical radiation fields might
also contribute to the IC emission. However, it is often very difficult to estimate the spectrum of the local radiation fields owing to
the poor knowledge of the environment. According to the interstellar radiation field model by \citet{2006ApJ...648L..29P} the contribution of IR and optical radiation fields should be negligible because of the large distance between SNR \velajr\ and the Galactic center ($\sim\!\!9$\,kpc).
Stars detected in the field of view of \velajr\ are not powerful enough to provide a strong radiation field and, moreover, the high uncertainty
on the determination of the distance to these stars makes it difficult to judge whether the stars are located in the proximity of the remnant or
not \citep{2010A&A...519A..86I}. Therefore, we decided to adopt CMB as the only radiation field responsible for the IC scattering of relativistic electrons.
The \gr\ emission is calculated according to \citet{1970RvMP...42..237B}.
The leptonic model (red dashed line in Fig.~\ref{fig:gamma_sed}, left) provides a good fit to the \FERMI-\HESS\ data with $\chi^2/\mathrm{NDF} = 13.8/13$ ($\chi^2$ probability 0.39).
The best-fit parameters are shown in Table~\ref{tab:ParamsParentParticles}.
The integration of
the electron spectrum above $100$\,GeV yields a total energy in electrons
$W_{\mathrm{e}} = (4.1 \pm 0.3\stat \pm 1.7\syst) \times 10^{47}$\,erg.

A magnetic field strength of $7\,\mu$G is needed to explain the observed \xr\ flux by synchrotron emission from the
obtained electron population (red dotted line in Fig.~\ref{fig:gamma_sed}, right). The modeled synchrotron spectrum does not reproduce well
the slope of the observed \xr\ spectrum within its statistical uncertainty, but both agree within the systematics of the \xr\ spectrum.
The same synchrotron emission overpredicts the radio flux. However, a decrease in the values of all parameters by $0.6\sigma$ of the
quadratically added statistical and systematic uncertainties allows us to accommodate the radio data within our leptonic model
(gray dotted line in Fig.~\ref{fig:gamma_sed}, right).

For such a low magnetic field, synchrotron cooling effects are negligible and do not have a significant impact on the energy spectrum of the
electron population. Indeed, following \citet{1970RvMP...42..237B} and adopting an SNR age of 3000 yr, the break energy above which the synchrotron losses become important is
\begin{equation}
  E_{\mathrm{b}} \simeq 90 \left(\frac{t_{\mathrm{age}}}{3000\,\mathrm{yr}}\right)^{-1} \left(\frac{B}{7\,\mu\mathrm{G}}\right)^{-2}\,\mathrm{TeV}, \nonumber
\end{equation}
which is considerably higher than the obtained electron cutoff energy.
This result suggests that the electron spectrum of \velajr\ may be limited by the age of the SNR \citep[see, e.g.,][]{reynolds98} rather than by radiative losses or that the SNR is considerably younger than 3000 yr.
A typical acceleration time for the electrons to reach the energy of 27\,TeV in the magnetic field of $7\,\mu$G for the shock compression
ratio of 4 is \citep{parizot06}
\begin{equation}
  t_{\mathrm{acc}} = 1.3 \eta \left( \frac{E_{\mathrm{cut},\,\mathrm{e}}}{27\,\mathrm{TeV}} \right) \left( \frac{B}{7\,\mu\mathrm{G}} \right)^{-1} \left( \frac{V_\mathrm{s}}{3000\,\mathrm{km\,s^{-1}}} \right)^{-2} \mathrm{kyr},\nonumber
\end{equation}
where $V_\mathrm{s}$ is the shock velocity and $\eta$ expresses the deviation of the diffusion coefficient from Bohm diffusion. However,
the magnetic field could have been higher in the past, in which case synchrotron losses can still play a major role.

The obtained value of the magnetic field strength is in very good agreement
with the values obtained in other leptonic models available in the
literature \citep[see, e.g.,][]{2013ApJ...767...20L}. This value represents the
average magnetic field across the remnant and does not exclude the
existence of regions with higher or lower magnetic fields; it is, however,
in conflict with the significantly amplified magnetic fields
derived for the filamentary structures in the NW rim under the hypothesis
that the filament width is limited by synchrotron cooling.
Nevertheless, the good agreement in the morphology of the \xr\ and \gr\ emissions,
as shown by the contours of Fig.~\ref{fig:skymaps} and the azimuthal profiles of Fig.~\ref{fig:az_prof},
still supports a leptonic nature of the \gr\ emission.
%

\subsubsection{Hadronic scenario}

In the hadronic scenario, we assume that the GeV-TeV \gr\ emission from \velajr\ is dominated by \grs\ produced in hadronic interactions.
The \gr\ emission is calculated according to \citet{2006PhRvD..74c4018K}.
The hadronic model (solid blue line in Fig.~\ref{fig:gamma_sed}, left) fits the \FERMI-\HESS\ data very well with $\chi^2/\mathrm{NDF} = 6.0/13$ ($\chi^2$ probability 0.95).
The best-fit parameters are shown in Table~\ref{tab:ParamsParentParticles}.
The integration of the proton spectrum above $1$\,GeV yields a total energy in protons
$W_{\mathrm{p}} = (7.1 \pm 0.3\stat \pm 1.9\syst) \times 10^{49} \left[\frac{n}{1\,\mathrm{cm}^{-3}}\right]^{-1}$\,erg, where $n$ is the density of the ambient medium.

The lack of thermal \xrs\ suggests a density of the ambient medium that is significantly lower than $1\,\mathrm{cm}^{-3}$ \citep{2001ApJ...548..814S}.
This, in turn, would require an unrealistically high estimate of the energy transferred to protons, comparable to or even higher than the total energy of
$\sim\!\!10^{51}$\,erg provided by the SN explosion. This problem, however, can be solved if the SNR is expanding in a very inhomogeneous clumpy environment
with compact dense clouds
as shown for \rxj\ \citep{2012ApJ...744...71I,2014MNRAS.445L..70G}. In this case the shock does not penetrate deep enough inside the clouds
to heat up the gas and generate thermal \xrs, but at the same time relativistic protons can penetrate inside to interact with the cloud material
and produce \gr\ emission. Indeed, \velajr\ exhibits a good correspondence between the TeV
\gr\ emission and the column density of interstellar medium proton distribution \citep[see, e.g.,][]{2013ASSP...34..249F}, which suggests
a hadronic origin of at least a portion of the observed \gr\ emission. However, modeling such a scenario is beyond the scope of this paper.

The proton spectral index obtained in the GeV-TeV fit ($1.83 \pm 0.02\stat \pm 0.11\syst$) is slightly harder than the value of 2.0 expected
in the diffusive shock acceleration for strong shocks with compression ratio of 4. However, from most of the SNRs detected at gamma-ray energies, an
even steeper spectrum of protons is inferred \citep[see, e.g.,][]{caprioli11}. The hard proton spectrum (index $1.8$) implied by the fit of the GeV-TeV emission from \velajr\
requires further theoretical investigation.
%

\subsection{Pulsar wind nebula around \psr}\label{interpretation:pwn}

The pulsar population studies by \citet{paper:hess_pwnpop}, and references therein, show that wind nebulae of energetic pulsars are very likely to be detectable in \grs\ at TeV energies. Since \psr\ falls into this category, it is likely that some of the emission attributed to \velajr\ could come from a possible TeV PWN associated with \psr.
The \HESS\ point spread function does not allow for a separation of the emissions due to the PWN and SNR.
Nevertheless, the flux measurements in the regions A--C from Table~\ref{tab:spec_morph_spectral_pars} can be interpreted as flux upper limits on a possible TeV PWN associated with \psr\ for different size assumptions of the PWN. Different
sizes of the putative TeV PWN were assumed owing to the lack of knowledge of its potential size at TeV energies. For pulsars older than $\sim\!\!10$\,kyr the size of the TeV PWN can be up to $100\!\!-\!\!1000$ times larger than
the size of the \xr\ PWN \citep{2010AIPC.1248...25K} owing to particle propagation up to large distances from the pulsar through diffusion/advection processes and/or proper motion of the pulsar itself.
In this context, the relatively small size of the \xr\ PWN associated with \psr\ (150\,arcsec, slightly smaller than the marker used to mark the pulsar position in the skymaps, i.e.,~in Fig.~\ref{fig:skymaps}, right) compared to the extension of \velajr\ may be the reason for the smaller enhancement seen in the azimuthal profile (cf.~Fig.~\ref{fig:az_prof}) in \xrs\ compared to \grs\ toward the direction of \psr.

Region B is the only region with spectral parameters deviating
more than $3\sigma$ ($3.5\sigma$ pretrials, $2.6\sigma$ post-trials)
from the spectrum of the NW rim, which might be suggestive of a PWN contribution to the flux
in this region. The insignificant deviation of the spectra of the regions A and C from the
rest of the remnant can be naturally explained by the sizes of these two regions.
On the one hand, the small size of region A results in
low statistics and hence high uncertainties in the spectral parameters. On the other hand, with its much larger size, region C comprises a
significant part of the SNR interior and other parts of the shell, leading to the domination of the emission from the SNR over
the emission from the PWN.
Therefore, the deviation of the spectrum of region B from the spectrum of the rest of the SNR not only
points to the contribution from the PWN but also constrains the size of the putative PWN. Moreover, the \gr\ spectrum of region
B deviates from the spectrum of the rest of the SNR in the same way as the \xr\ spectrum of the PWN differs from the
spectrum of the SNR, which further supports the hypothesis of significant contribution of the PWN emission to the \gr\ flux
in region B. The \xr\ spectrum of the PWN is significantly harder than the spectrum of the SNR \citep{2013A&A...551A...7A},
which agrees well with an indication of increase of the cutoff energy of the \gr\ spectrum of region B with respect to the rest of the remnant.
However, the significance of the increase of the cutoff energy in region B as well as of the
existence of the cutoff feature is low. Setting the observed flux from region B as an upper limit on the flux from the PWN
one can estimate the upper limit on the efficiency of the conversion of the pulsar spin-down luminosity to the \gr\ emission.
The upper limit on the TeV efficiency defined as the ratio of the $1\!\!-\!\!10$\,TeV PWN luminosity to the spin-down power is $0.75 \times 10^{-3}$,
using for the distance to the pulsar the upper limit of 900\,pc from \citet{2013A&A...551A...7A}.
This value of the \gr\ efficiency is compatible with that of other PWNe detected
at TeV energies \citep{paper:hess_pwnpop}.

The \gr\ emission from the PWN might provide a substantial contribution to the overall flux
from \velajr. Adopting the size of region B as a potential size of the putative TeV PWN, this
contribution can be as large as $\sim\!\!8\%$ of the total \gr\ flux from the SNR, considering the flux
level measured for region B.
However, the \gr\ emission from the remnant
also contributes to the flux from region B and hence only a fraction of the \gr\ emission
detected in that region is from the PWN. The azimuthal profile (cf.~Fig.~\ref{fig:az_prof}) exhibits
approximately double the flux at the position coincident with the \xr\ PWN compared to the regions around
it. Therefore, a more plausible prediction for the fraction of the \gr\ emission from the PWN in the
overall flux from \velajr\ is about $4\%$. This was not taken into account in the modeling of the SNR presented
above, assuming that all the detected \gr\ flux is coming from \velajr.
However, if the contribution of the PWN is indeed
only around $4\%$ of the overall flux, it would be well covered by the systematic errors estimated for the GeV and TeV data. Even an $8\%$ contribution would be covered by the $25\%$ systematic error assigned to the overall flux.
%

\section{Conclusions}\label{sec:conclusions}

The analysis of an enlarged \HESS\ data set, using approximately double the observation time compared to previous publications, comes to the following conclusions:
\begin{enumerate}
  \item A revised flux measurement makes \velajr\ the brightest steady source in the sky above 1\,TeV with $F(>\!\!1 \UNITS{TeV}) = (23.4 \pm 0.7\stat \pm 4.9\syst) \times 10^{-12} \UNITS{cm^{-2} s^{-1}}$; this flux is $\sim\!\!13\%$ larger than the flux of the Crab nebula in the same energy range.
  \item The energy spectrum of \velajr\ is clearly curved and has an exponential cutoff at $E_\mathrm{cut} = (6.7 \pm 1.2\stat \pm 1.2\syst) \UNITS{TeV}$. The determination of the cutoff helps to characterize the parent particle population better.
  \item The new TeV spectrum connects well with the \FERMI\ measurement at GeV energies without the need for a spectral break as in previous publications. This smooth connection of the GeV and TeV spectra together with the well-defined cutoff allows us to determine directly the characteristics of the parent particle population in both leptonic and hadronic scenarios.
  \item The study of the spatially resolved spectroscopy reveals no clear spectral variation across the SNR, suggesting that the parent particle population can be assumed to be the same throughout the remnant, which in turn indicates that the conditions for particle acceleration, i.e.,~properties of the SNR shock and ambient medium, are similar everywhere.
  \item The enhancement detected toward \psr\ suggests that some of the emission might come from a possible TeV PWN associated with the pulsar.
The contribution is estimated to be less than 8\%.
\end{enumerate}

Deeper \HESS\ observations of \velajr\ thus provide a significant improvement of statistics revealing the existence of a
cutoff in the TeV spectrum, which in turn considerably improves the characterization of the parent particle population
and allows, for the first time, the estimation of the uncertainties of its parameters. Both leptonic and hadronic models remain plausible as both models provide a good fit of the data.
Larger statistics at high energies (i.e.,~above $10\,$TeV), which provide a better characterization of the spectral cutoff,
could give more insight into the nature of the parent particle population for this object. Such data could be obtained with
even deeper \HESS\ observations and ultimately from the future CTA \citep{CtaDesignConcepts} observatory.
In addition, larger statistics and better angular resolution should help in the search for clear spectral variations across the \velajr\ region, such as a softening of the spectrum, when moving away from \psr.
Moreover, smaller errors on the spectral parameters would
also allow the establishment of spectral differences between region B and the NW rim. All this would help to
separate the contributions from the PWN and SNR.
%

\begin{acknowledgements}
The support of the Namibian authorities and the University of Namibia in facilitating the construction and operation of \HESS\ is gratefully acknowledged,
as is the support by the German Ministry for Education and Research (BMBF), the Max Planck Society, the German Research Foundation (DFG), the
French Ministry for Research, the CNRS-IN2P3 and the Astroparticle Interdisciplinary Programme of the CNRS, the U.K.~Science and Technology
Facilities Council (STFC), the IPNP of the Charles University, the Czech Science Foundation, the Polish Ministry of Science and Higher Education,
the South African Department of Science and Technology and National Research Foundation, the University of Namibia,
the Innsbruck University, the Austrian Science Fund (FWF), and the Austrian Federal Ministry for Science, Research and Economy,
and by the University of Adelaide and the Australian Research Council. We appreciate the excellent work of the technical support staff
in Berlin, Durham, Hamburg, Heidelberg, Palaiseau, Paris, Saclay, and in Namibia in the construction and operation of the equipment.
This work benefitted from services provided by the \HESS\ Virtual Organisation, supported by the national resource providers of the EGI Federation.
\end{acknowledgements}
%

\bibliographystyle{aa} 
\bibliography{bibliography} 

\begin{thebibliography}{48}
\expandafter\ifx\csname natexlab\endcsname\relax\def\natexlab#1{#1}\fi

\bibitem[{{Abramowski} {et~al.}(2012){Abramowski}, {Acero}, {Aharonian},
  {Akhperjanian}, {Anton}, {Balenderan}, {Balzer}, {Barnacka}, {Becherini},
  {Becker Tjus}, {Bernl{\"o}hr}, {Birsin}, {Biteau}, {Bochow}, {Boisson},
  {Bolmont}, {Bordas}, {Brucker}, {Brun}, {Brun}, {Bulik}, {Carrigan},
  {Casanova}, {Cerruti}, {Chadwick}, {Charbonnier}, {Chaves}, {Cheesebrough},
  {Cologna}, {Conrad}, {Couturier}, {Dalton}, {Daniel}, {Davids}, {Degrange},
  {Deil}, {deWilt}, {Dickinson}, {Djannati-Ata{\"i}}, {Domainko}, {Drury},
  {Dubois}, {Dubus}, {Dutson}, {Dyks}, {Dyrda}, {Egberts}, {Eger}, {Espigat},
  {Fallon}, {Farnier}, {Fegan}, {Feinstein}, {Fernandes}, {Fernandez},
  {Fiasson}, {Fontaine}, {F{\"o}rster}, {F{\"u}{\ss}ling}, {Gajdus}, {Gallant},
  {Garrigoux}, {Gast}, {Giebels}, {Glicenstein}, {Gl{\"u}ck}, {G{\"o}ring},
  {Grondin}, {H{\"a}ffner}, {Hague}, {Hahn}, {Hampf}, {Harris}, {Heinz},
  {Heinzelmann}, {Henri}, {Hermann}, {Hillert}, {Hinton}, {Hofmann},
  {Hofverberg}, {Holler}, {Horns}, {Jacholkowska}, {Jahn}, {Jamrozy}, {Jung},
  {Kastendieck}, {Katarzy{\'n}ski}, {Katz}, {Kaufmann}, {Kh{\'e}lifi},
  {Klochkov}, {Klu{\'z}niak}, {Kneiske}, {Komin}, {Kosack}, {Kossakowski},
  {Krayzel}, {Kr{\"u}ger}, {Laffon}, {Lamanna}, {Lenain}, {Lennarz}, {Lohse},
  {Lopatin}, {Lu}, {Marandon}, {Marcowith}, {Masbou}, {Maurin}, {Maxted},
  {Mayer}, {McComb}, {Medina}, {M{\'e}hault}, {Menzler}, {Moderski}, {Mohamed},
  {Moulin}, {Naumann}, {Naumann-Godo}, {de Naurois}, {Nedbal}, {Nguyen},
  {Niemiec}, {Nolan}, {Ohm}, {de O{\~n}a Wilhelmi}, {Opitz}, {Ostrowski},
  {Oya}, {Panter}, {Parsons}, {Paz Arribas}, {Pekeur}, {Pelletier}, {Perez},
  {Petrucci}, {Peyaud}, {Pita}, {P{\"u}hlhofer}, {Punch}, {Quirrenbach},
  {Raue}, {Reimer}, {Reimer}, {Renaud}, {de los Reyes}, {Rieger}, {Ripken},
  {Rob}, {Rosier-Lees}, {Rowell}, {Rudak}, {Rulten}, {Sahakian}, {Sanchez},
  {Santangelo}, {Schlickeiser}, {Schulz}, {Schwanke}, {Schwarzburg},
  {Schwemmer}, {Sheidaei}, {Skilton}, {Sol}, {Spengler}, {Stawarz},
  {Steenkamp}, {Stegmann}, {Stinzing}, {Stycz}, {Sushch}, {Szostek},
  {Tavernet}, {Terrier}, {Tluczykont}, {Trichard}, {Valerius}, {van Eldik},
  {Vasileiadis}, {Venter}, {Viana}, {Vincent}, {V{\"o}lk}, {Volpe}, {Vorobiov},
  {Vorster}, {Wagner}, {Ward}, {White}, {Wierzcholska}, {Wouters}, {Zacharias},
  {Zajczyk}, {Zdziarski}, {Zech}, \& {Zechlin}}]{velax}
{Abramowski}, A., {Acero}, F., {Aharonian}, F., {et~al.} 2012, \aap, 548, A38

\bibitem[{{Acero} {et~al.}(2015){Acero}, {Ackermann}, {Ajello}, {Albert},
  {Atwood}, {Axelsson}, {Baldini}, {Ballet}, {Barbiellini}, {Bastieri},
  {Belfiore}, {Bellazzini}, {Bissaldi}, {Blandford}, {Bloom}, {Bogart},
  {Bonino}, {Bottacini}, {Bregeon}, {Britto}, {Bruel}, {Buehler}, {Burnett},
  {Buson}, {Caliandro}, {Cameron}, {Caputo}, {Caragiulo}, {Caraveo},
  {Casandjian}, {Cavazzuti}, {Charles}, {Chaves}, {Chekhtman}, {Cheung},
  {Chiang}, {Chiaro}, {Ciprini}, {Claus}, {Cohen-Tanugi}, {Cominsky}, {Conrad},
  {Cutini}, {D'Ammando}, {de Angelis}, {DeKlotz}, {de Palma}, {Desiante},
  {Digel}, {Di Venere}, {Drell}, {Dubois}, {Dumora}, {Favuzzi}, {Fegan},
  {Ferrara}, {Finke}, {Franckowiak}, {Fukazawa}, {Funk}, {Fusco}, {Gargano},
  {Gasparrini}, {Giebels}, {Giglietto}, {Giommi}, {Giordano}, {Giroletti},
  {Glanzman}, {Godfrey}, {Grenier}, {Grondin}, {Grove}, {Guillemot}, {Guiriec},
  {Hadasch}, {Harding}, {Hays}, {Hewitt}, {Hill}, {Horan}, {Iafrate}, {Jogler},
  {J{\'o}hannesson}, {Johnson}, {Johnson}, {Johnson}, {Johnson}, {Kamae},
  {Kataoka}, {Katsuta}, {Kuss}, {La Mura}, {Landriu}, {Larsson}, {Latronico},
  {Lemoine-Goumard}, {Li}, {Li}, {Longo}, {Loparco}, {Lott}, {Lovellette},
  {Lubrano}, {Madejski}, {Massaro}, {Mayer}, {Mazziotta}, {McEnery},
  {Michelson}, {Mirabal}, {Mizuno}, {Moiseev}, {Mongelli}, {Monzani},
  {Morselli}, {Moskalenko}, {Murgia}, {Nuss}, {Ohno}, {Ohsugi}, {Omodei},
  {Orienti}, {Orlando}, {Ormes}, {Paneque}, {Panetta}, {Perkins},
  {Pesce-Rollins}, {Piron}, {Pivato}, {Porter}, {Racusin}, {Rando}, {Razzano},
  {Razzaque}, {Reimer}, {Reimer}, {Reposeur}, {Rochester}, {Romani},
  {Salvetti}, {S{\'a}nchez-Conde}, {Saz Parkinson}, {Schulz}, {Siskind},
  {Smith}, {Spada}, {Spandre}, {Spinelli}, {Stephens}, {Strong}, {Suson},
  {Takahashi}, {Takahashi}, {Tanaka}, {Thayer}, {Thayer}, {Thompson},
  {Tibaldo}, {Tibolla}, {Torres}, {Torresi}, {Tosti}, {Troja}, {Van Klaveren},
  {Vianello}, {Winer}, {Wood}, {Wood}, {Zimmer}, \& {Fermi-LAT
  Collaboration}}]{Fermi3FGL}
{Acero}, F., {Ackermann}, M., {Ajello}, M., {et~al.} 2015, \apjs, 218, 23

\bibitem[{{Acero} {et~al.}(2013){Acero}, {Gallant}, {Ballet}, {Renaud}, \&
  {Terrier}}]{2013A&A...551A...7A}
{Acero}, F., {Gallant}, Y., {Ballet}, J., {Renaud}, M., \& {Terrier}, R. 2013,
  \aap, 551, A7

\bibitem[{{Actis} {et~al.}(2011){Actis}, {Agnetta}, {Aharonian},
  {Akhperjanian}, {Aleksi{\'c}}, {Aliu}, {Allan}, {Allekotte}, {Antico},
  {Antonelli}, \& et~al.}]{CtaDesignConcepts}
{Actis}, M., {Agnetta}, G., {Aharonian}, F., {et~al.} 2011, Experimental
  Astronomy, 32, 193

\bibitem[{{Aharonian} {et~al.}(2006){Aharonian}, {Akhperjanian}, {Bazer-Bachi},
  {Beilicke}, {Benbow}, {Berge}, {Bernl{\"o}hr}, {Boisson}, {Bolz}, {Borrel},
  {Braun}, {Breitling}, {Brown}, {B{\"u}hler}, {B{\"u}sching}, {Carrigan},
  {Chadwick}, {Chounet}, {Cornils}, {Costamante}, {Degrange}, {Dickinson},
  {Djannati-Ata{\"i}}, {O'C.~Drury}, {Dubus}, {Egberts}, {Emmanoulopoulos},
  {Espigat}, {Feinstein}, {Ferrero}, {Fiasson}, {Fontaine}, {Funk}, {Funk},
  {Gallant}, {Giebels}, {Glicenstein}, {Goret}, {Hadjichristidis}, {Hauser},
  {Hauser}, {Heinzelmann}, {Henri}, {Hermann}, {Hinton}, {Hofmann}, {Holleran},
  {Horns}, {Jacholkowska}, {de Jager}, {Kh{\'e}lifi}, {Komin}, {Konopelko},
  {Kosack}, {Latham}, {Le Gallou}, {Lemi{\`e}re}, {Lemoine-Goumard}, {Lohse},
  {Martin}, {Martineau-Huynh}, {Marcowith}, {Masterson}, {McComb}, {de
  Naurois}, {Nedbal}, {Nolan}, {Noutsos}, {Orford}, {Osborne}, {Ouchrif},
  {Panter}, {Pelletier}, {Pita}, {P{\"u}hlhofer}, {Punch}, {Raubenheimer},
  {Raue}, {Rayner}, {Reimer}, {Reimer}, {Ripken}, {Rob}, {Rolland}, {Rowell},
  {Sahakian}, {Saug{\'e}}, {Schlenker}, {Schlickeiser}, {Schwanke}, {Sol},
  {Spangler}, {Spanier}, {Steenkamp}, {Stegmann}, {Superina}, {Tavernet},
  {Terrier}, {Th{\'e}oret}, {Tluczykont}, {van Eldik}, {Vasileiadis}, {Venter},
  {Vincent}, {V{\"o}lk}, {Wagner}, \& {Ward}}]{paper:hess_crab}
{Aharonian}, F., {Akhperjanian}, A.~G., {Bazer-Bachi}, A.~R., {et~al.} 2006,
  \aap, 457, 899

\bibitem[{{Aharonian} {et~al.}(2005){Aharonian}, {Akhperjanian}, {Bazer-Bachi},
  {Beilicke}, {Benbow}, {Berge}, {Bernl{\"o}hr}, {Boisson}, {Bolz}, {Borrel},
  {Braun}, {Breitling}, {Brown}, {Chadwick}, {Chounet}, {Cornils},
  {Costamante}, {Degrange}, {Dickinson}, {Djannati-Ata{\"i}}, {O'C.~Drury},
  {Dubus}, {Emmanoulopoulos}, {Espigat}, {Feinstein}, {Fontaine}, {Fuchs},
  {Funk}, {Gallant}, {Giebels}, {Gillessen}, {Glicenstein}, {Goret},
  {Hadjichristidis}, {Hauser}, {Heinzelmann}, {Henri}, {Hermann}, {Hinton},
  {Hofmann}, {Holleran}, {Horns}, {Jacholkowska}, {de Jager}, {Kh{\'e}lifi},
  {Komin}, {Konopelko}, {Latham}, {Le Gallou}, {Lemi{\`e}re},
  {Lemoine-Goumard}, {Leroy}, {Lohse}, {Martin}, {Martineau-Huynh},
  {Marcowith}, {Masterson}, {McComb}, {de Naurois}, {Nolan}, {Noutsos},
  {Orford}, {Osborne}, {Ouchrif}, {Panter}, {Pelletier}, {Pita},
  {P{\"u}hlhofer}, {Punch}, {Raubenheimer}, {Raue}, {Raux}, {Rayner}, {Reimer},
  {Reimer}, {Ripken}, {Rob}, {Rolland}, {Rowell}, {Sahakian}, {Saug{\'e}},
  {Schlenker}, {Schlickeiser}, {Schuster}, {Schwanke}, {Siewert}, {Sol},
  {Spangler}, {Steenkamp}, {Stegmann}, {Tavernet}, {Terrier}, {Th{\'e}oret},
  {Tluczykont}, {Vasileiadis}, {Venter}, {Vincent}, {V{\"o}lk}, \&
  {Wagner}}]{paper:vjr_hess_paper1}
{Aharonian}, F., {Akhperjanian}, A.~G., {Bazer-Bachi}, A.~R., {et~al.} 2005,
  \aap, 437, L7

\bibitem[{{Aharonian} {et~al.}(2007){Aharonian}, {Akhperjanian}, {Bazer-Bachi},
  {Beilicke}, {Benbow}, {Berge}, {Bernl{\"o}hr}, {Boisson}, {Bolz}, {Borrel},
  {Braun}, {Brown}, {B{\"u}hler}, {B{\"u}sching}, {Carrigan}, {Chadwick},
  {Chounet}, {Coignet}, {Cornils}, {Costamante}, {Degrange}, {Dickinson},
  {Djannati-Ata{\"i}}, {Drury}, {Dubus}, {Egberts}, {Emmanoulopoulos},
  {Espigat}, {Feinstein}, {Ferrero}, {Fiasson}, {Filipovic}, {Fontaine},
  {Fukui}, {Funk}, {Funk}, {F{\"u}{\ss}ling}, {Gallant}, {Giebels},
  {Glicenstein}, {Goret}, {Hadjichristidis}, {Hauser}, {Hauser}, {Heinzelmann},
  {Henri}, {Hermann}, {Hinton}, {Hiraga}, {Hoffmann}, {Hofmann}, {Holleran},
  {Hoppe}, {Horns}, {Ishisaki}, {Jacholkowska}, {de Jager}, {Kendziorra},
  {Kerschhaggl}, {Kh{\'e}lifi}, {Komin}, {Konopelko}, {Kosack}, {Lamanna},
  {Latham}, {Le Gallou}, {Lemi{\`e}re}, {Lemoine-Goumard}, {Lohse}, {Martin},
  {Martineau-Huynh}, {Marcowith}, {Masterson}, {Maurin}, {McComb}, {Moulin},
  {Moriguchi}, {de Naurois}, {Nedbal}, {Nolan}, {Noutsos}, {Orford}, {Osborne},
  {Ouchrif}, {Panter}, {Pelletier}, {Pita}, {P{\"u}hlhofer}, {Punch},
  {Ranchon}, {Raubenheimer}, {Raue}, {Rayner}, {Reimer}, {Ripken}, {Rob},
  {Rolland}, {Rosier-Lees}, {Rowell}, {Sahakian}, {Santangelo}, {Saug{\'e}},
  {Schlenker}, {Schlickeiser}, {Schr{\"o}der}, {Schwanke}, {Schwarzburg},
  {Schwemmer}, {Shalchi}, {Sol}, {Spangler}, {Spanier}, {Steenkamp},
  {Stegmann}, {Superina}, {Tam}, {Tavernet}, {Terrier}, {Tluczykont}, {van
  Eldik}, {Vasileiadis}, {Venter}, {Vialle}, {Vincent}, {V{\"o}lk}, {Wagner},
  \& {Ward}}]{paper:vjr_hess_paper2}
{Aharonian}, F., {Akhperjanian}, A.~G., {Bazer-Bachi}, A.~R., {et~al.} 2007,
  \apj, 661, 236

\bibitem[{{Allen} {et~al.}(2015){Allen}, {Chow}, {DeLaney}, {Filipovi{\'c}},
  {Houck}, {Pannuti}, \& {Stage}}]{2015ApJ...798...82A}
{Allen}, G.~E., {Chow}, K., {DeLaney}, T., {et~al.} 2015, \apj, 798, 82

\bibitem[{{Aschenbach}(1998)}]{1998Natur.396..141A}
{Aschenbach}, B. 1998, \nat, 396, 141

\bibitem[{{Aschenbach} {et~al.}(1999){Aschenbach}, {Iyudin}, \&
  {Sch{\"o}nfelder}}]{1999A&A...350..997A}
{Aschenbach}, B., {Iyudin}, A.~F., \& {Sch{\"o}nfelder}, V. 1999, \aap, 350,
  997

\bibitem[{{Bamba} {et~al.}(2005){Bamba}, {Yamazaki}, \&
  {Hiraga}}]{2005ApJ...632..294B}
{Bamba}, A., {Yamazaki}, R., \& {Hiraga}, J.~S. 2005, \apj, 632, 294

\bibitem[{{Berezhko} {et~al.}(2009){Berezhko}, {P{\"u}hlhofer}, \&
  {V{\"o}lk}}]{2009A&A...505..641B}
{Berezhko}, E.~G., {P{\"u}hlhofer}, G., \& {V{\"o}lk}, H.~J. 2009, \aap, 505,
  641

\bibitem[{{Berge} {et~al.}(2007){Berge}, {Funk}, \& {Hinton}}]{paper:bg}
{Berge}, D., {Funk}, S., \& {Hinton}, J. 2007, \aap, 466, 1219

\bibitem[{{Blumenthal} \& {Gould}(1970)}]{1970RvMP...42..237B}
{Blumenthal}, G.~R. \& {Gould}, R.~J. 1970, Reviews of Modern Physics, 42, 237

\bibitem[{{Caprioli}(2011)}]{caprioli11}
{Caprioli}, D. 2011, \jcap, 5, 026

\bibitem[{{Duncan} \& {Green}(2000)}]{paper:vjr_radio_points}
{Duncan}, A.~R. \& {Green}, D.~A. 2000, \aap, 364, 732

\bibitem[{{Fukui}(2013)}]{2013ASSP...34..249F}
{Fukui}, Y. 2013, in Astrophysics and Space Science Proceedings, Vol.~34,
  Cosmic Rays in Star-Forming Environments, ed. D.~F. {Torres} \& O.~{Reimer},
  249

\bibitem[{{Gabici} \& {Aharonian}(2014)}]{2014MNRAS.445L..70G}
{Gabici}, S. \& {Aharonian}, F.~A. 2014, \mnras, 445, L70

\bibitem[{{Green}(2009)}]{cat:green_snr}
{Green}, D.~A. 2009, Bulletin of the Astronomical Society of India, 37, 45

\bibitem[{{H.~E.~S.~S. Collaboration} {et~al.}(2018{\natexlab{a}}){H.~E.~S.~S.
  Collaboration}, {Abdalla}, {Abramowski}, {Aharonian}, {Ait Benkhali},
  {Akhperjanian}, {Andersson}, {Ang{\"u}ner}, {Arrieta}, {Aubert}, {Backes},
  {Balzer}, {Barnard}, {Becherini}, {Becker Tjus}, {Berge}, {Bernhard},
  {Bernl{\"o}hr}, {Blackwell}, {B{\"o}ttcher}, {Boisson}, {Bolmont}, {Bordas},
  {Bregeon}, {Brun}, {Brun}, {Bryan}, {Bulik}, {Capasso}, {Carr}, {Casanova},
  {Cerruti}, {Chakraborty}, {Chalme-Calvet}, {Chaves}, {Chen}, {Chevalier},
  {Chr{\'e}tien}, {Colafrancesco}, {Cologna}, {Condon}, {Conrad}, {Cui},
  {Davids}, {Decock}, {Degrange}, {Deil}, {Devin}, {deWilt}, {Dirson},
  {Djannati-Ata{\"\i}}, {Domainko}, {Donath}, {Drury}, {Dubus}, {Dutson},
  {Dyks}, {Edwards}, {Egberts}, {Eger}, {Ernenwein}, {Eschbach}, {Farnier},
  {Fegan}, {Fernandes}, {Fiasson}, {Fontaine}, {F{\"o}rster}, {Fukuyama},
  {Funk}, {F{\"u}{\ss}ling}, {Gabici}, {Gajdus}, {Gallant}, {Garrigoux},
  {Giavitto}, {Giebels}, {Glicenstein}, {Gottschall}, {Goyal}, {Grondin},
  {Hadasch}, {Hahn}, {Haupt}, {Hawkes}, {Heinzelmann}, {Henri}, {Hermann},
  {Hervet}, {Hinton}, {Hofmann}, {Hoischen}, {Holler}, {Horns}, {Ivascenko},
  {Jacholkowska}, {Jamrozy}, {Janiak}, {Jankowsky}, {Jankowsky}, {Jingo},
  {Jogler}, {Jouvin}, {Jung- Richardt}, {Kastendieck}, {Katarzy{\'n}ski},
  {Katz}, {Kerszberg}, {Kh{\'e}lifi}, {Kieffer}, {King}, {Klepser}, {Klochkov},
  {Klu{\'z}niak}, {Kolitzus}, {Komin}, {Kosack}, {Krakau}, {Kraus}, {Krayzel},
  {Kr{\"u}ger}, {Laffon}, {Lamanna}, {Lau}, {Lees}, {Lefaucheur}, {Lefranc},
  {Lemi{\`e}re}, {Lemoine-Goumard}, {Lenain}, {Leser}, {Lohse}, {Lorentz},
  {Liu}, {L{\'o}pez-Coto}, {Lypova}, {Marandon}, {Marcowith}, {Mariaud},
  {Marx}, {Maurin}, {Maxted}, {Mayer}, {Meintjes}, {Meyer}, {Mitchell},
  {Moderski}, {Mohamed}, {Mohrmann}, {Mor{\r{a}}}, {Moulin}, {Murach}, {de
  Naurois}, {Niederwanger}, {Niemiec}, {Oakes}, {O'Brien}, {Odaka}, {{\"O}ttl},
  {Ohm}, {Ostrowski}, {Oya}, {Padovani}, {Panter}, {Parsons}, {Pekeur},
  {Pelletier}, {Perennes}, {Petrucci}, {Peyaud}, {Piel}, {Pita}, {Poon},
  {Prokhorov}, {Prokoph}, {P{\"u}hlhofer}, {Punch}, {Quirrenbach}, {Raab},
  {Reimer}, {Reimer}, {Renaud}, {de los Reyes}, {Rieger}, {Romoli},
  {Rosier-Lees}, {Rowell}, {Rudak}, {Rulten}, {Sahakian}, {Salek}, {Sanchez},
  {Santangelo}, {Sasaki}, {Schlickeiser}, {Sch{\"u}ssler}, {Schulz},
  {Schwanke}, {Schwemmer}, {Settimo}, {Seyffert}, {Shafi}, {Shilon}, {Simoni},
  {Sol}, {Spanier}, {Spengler}, {Spies}, {Stawarz}, {Steenkamp}, {Stegmann},
  {Stinzing}, {Stycz}, {Sushch}, {Takahashi}, {Tavernet}, {Tavernier},
  {Taylor}, {Terrier}, {Tibaldo}, {Tiziani}, {Tluczykont}, {Trichard}, {Tuffs},
  {Uchiyama}, {van der Walt}, {van Eldik}, {van Rensburg}, {van Soelen},
  {Vasileiadis}, {Veh}, {Venter}, {Viana}, {Vincent}, {Vink}, {Voisin},
  {V{\"o}lk}, {Volpe}, {Vuillaume}, {Wadiasingh}, {Wagner}, {Wagner}, {Wagner},
  {White}, {Wierzcholska}, {Willmann}, {W{\"o}rnlein}, {Wouters}, {Yang},
  {Zabalza}, {Zaborov}, {Zacharias}, {Zdziarski}, {Zech}, {Zefi}, {Ziegler}, \&
  {{\.Z}ywucka}}]{RXJ1713Forth}
{H.~E.~S.~S. Collaboration}, {Abdalla}, H., {Abramowski}, A., {et~al.}
  2018{\natexlab{a}}, \aap, 612

\bibitem[{{H.~E.~S.~S. Collaboration} {et~al.}(2018{\natexlab{b}}){H.~E.~S.~S.
  Collaboration}, {Abdalla}, {Abramowski}, {Aharonian}, {Ait Benkhali},
  {Akhperjanian}, {Andersson}, {Ang{\"u}ner}, {Arrieta}, {Aubert}, {Backes},
  {Balzer}, {Barnard}, {Becherini}, {Becker Tjus}, {Berge}, {Bernhard},
  {Bernl{\"o}hr}, {Blackwell}, {B{\"o}ttcher}, {Boisson}, {Bolmont}, {Bordas},
  {Bregeon}, {Brun}, {Brun}, {Bryan}, {Bulik}, {Capasso}, {Carr}, {Carrigan},
  {Casanova}, {Cerruti}, {Chakraborty}, {Chalme-Calvet}, {Chaves}, {Chen},
  {Chevalier}, {Chr{\'e}tien}, {Colafrancesco}, {Cologna}, {Condon}, {Conrad},
  {Couturier}, {Cui}, {Davids}, {Degrange}, {Deil}, {Devin}, {deWilt},
  {Dirson}, {Djannati- Ata{\"\i}}, {Domainko}, {Donath}, {Drury}, {Dubus},
  {Dutson}, {Dyks}, {Edwards}, {Egberts}, {Eger}, {Ernenwein}, {Eschbach},
  {Farnier}, {Fegan}, {Fernandes}, {Fiasson}, {Fontaine}, {F{\"o}rster},
  {Funk}, {F{\"u}{\ss}ling}, {Gabici}, {Gajdus}, {Gallant}, {Garrigoux},
  {Giavitto}, {Giebels}, {Glicenstein}, {Gottschall}, {Goyal}, {Grondin},
  {Hadasch}, {Hahn}, {Haupt}, {Hawkes}, {Heinzelmann}, {Henri}, {Hermann},
  {Hervet}, {Hillert}, {Hinton}, {Hofmann}, {Hoischen}, {Holler}, {Horns},
  {Ivascenko}, {Jacholkowska}, {Jamrozy}, {Janiak}, {Jankowsky}, {Jankowsky},
  {Jingo}, {Jogler}, {Jouvin}, {Jung-Richardt}, {Kastendieck},
  {Katarzy{\'n}ski}, {Katz}, {Kerszberg}, {Kh{\'e}lifi}, {Kieffer}, {King},
  {Klepser}, {Klochkov}, {Klu{\'z}niak}, {Kolitzus}, {Komin}, {Kosack},
  {Krakau}, {Kraus}, {Krayzel}, {Kr{\"u}ger}, {Laffon}, {Lamanna}, {Lau},
  {Lees}, {Lefaucheur}, {Lefranc}, {Lemi{\`e}re}, {Lemoine-Goumard}, {Lenain},
  {Leser}, {Lohse}, {Lorentz}, {Liu}, {L{\'o}pez- Coto}, {Lypova}, {Marandon},
  {Marcowith}, {Mariaud}, {Marx}, {Maurin}, {Maxted}, {Mayer}, {Meintjes},
  {Meyer}, {Mitchell}, {Moderski}, {Mohamed}, {Mohrmann}, {Mor{\r{a}}},
  {Moulin}, {Murach}, {de Naurois}, {Niederwanger}, {Niemiec}, {Oakes},
  {O'Brien}, {Odaka}, {{\"O}ttl}, {Ohm}, {de O{\~n}a Wilhelmi}, {Ostrowski},
  {Oya}, {Padovani}, {Panter}, {Parsons}, {Paz Arribas}, {Pekeur}, {Pelletier},
  {Perennes}, {Petrucci}, {Peyaud}, {Pita}, {Poon}, {Prokhorov}, {Prokoph},
  {P{\"u}hlhofer}, {Punch}, {Quirrenbach}, {Raab}, {Reimer}, {Reimer},
  {Renaud}, {de los Reyes}, {Rieger}, {Romoli}, {Rosier-Lees}, {Rowell},
  {Rudak}, {Rulten}, {Sahakian}, {Salek}, {Sanchez}, {Santangelo}, {Sasaki},
  {Schlickeiser}, {Sch{\"u}ssler}, {Schulz}, {Schwanke}, {Schwemmer},
  {Settimo}, {Seyffert}, {Shafi}, {Shilon}, {Simoni}, {Sol}, {Spanier},
  {Spengler}, {Spies}, {Stawarz}, {Steenkamp}, {Stegmann}, {Stinzing}, {Stycz},
  {Sushch}, {Tavernet}, {Tavernier}, {Taylor}, {Terrier}, {Tibaldo}, {Tiziani},
  {Tluczykont}, {Trichard}, {Tuffs}, {Uchiyama}, {Valerius}, {van der Walt},
  {van Eldik}, {van Soelen}, {Vasileiadis}, {Veh}, {Venter}, {Viana},
  {Vincent}, {Vink}, {Voisin}, {V{\"o}lk}, {Vuillaume}, {Wadiasingh}, {Wagner},
  {Wagner}, {Wagner}, {White}, {Wierzcholska}, {Willmann}, {W{\"o}rnlein},
  {Wouters}, {Yang}, {Zabalza}, {Zaborov}, {Zacharias}, {Zdziarski}, {Zech},
  {Zefi}, {Ziegler}, \& {{\.Z}ywucka}}]{paper:hess_pwnpop}
{H.~E.~S.~S. Collaboration}, {Abdalla}, H., {Abramowski}, A., {et~al.}
  2018{\natexlab{b}}, \aap, 612

\bibitem[{{Inoue} {et~al.}(2012){Inoue}, {Yamazaki}, {Inutsuka}, \&
  {Fukui}}]{2012ApJ...744...71I}
{Inoue}, T., {Yamazaki}, R., {Inutsuka}, S., \& {Fukui}, Y. 2012, \apj, 744, 71

\bibitem[{{Iyudin} {et~al.}(2010){Iyudin}, {Pakhomov}, {Chugai}, {Greiner},
  {Axelsson}, {Larsson}, \& {Ryabchikova}}]{2010A&A...519A..86I}
{Iyudin}, A.~F., {Pakhomov}, Y.~V., {Chugai}, N.~N., {et~al.} 2010, \aap, 519,
  A86

\bibitem[{{Kargaltsev} \& {Pavlov}(2010)}]{2010AIPC.1248...25K}
{Kargaltsev}, O. \& {Pavlov}, G.~G. 2010, X-ray Astronomy 2009; Present Status,
  Multi-Wavelength Approach and Future Perspectives, 1248, 25

\bibitem[{{Katagiri} {et~al.}(2005){Katagiri}, {Enomoto}, {Ksenofontov},
  {Mori}, {Adachi}, {Asahara}, {Bicknell}, {Clay}, {Doi}, {Edwards}, {Gunji},
  {Hara}, {Hara}, {Hattori}, {Hayashi}, {Itoh}, {Kabuki}, {Kajino}, {Kawachi},
  {Kifune}, {Kiuchi}, {Kubo}, {Kurihara}, {Kurosaka}, {Kushida}, {Matsubara},
  {Miyashita}, {Mizumoto}, {Muraishi}, {Muraki}, {Naito}, {Nakamori}, {Nakase},
  {Nishida}, {Nishijima}, {Ohishi}, {Okumura}, {Patterson}, {Protheroe},
  {Sakamoto}, {Sakamoto}, {Swaby}, {Tanimori}, {Tanimura}, {Thornton},
  {Tsuchiya}, {Watanabe}, {Yamaoka}, {Yanagita}, {Yoshida}, \&
  {Yoshikoshi}}]{paper:vjr_cangaroo}
{Katagiri}, H., {Enomoto}, R., {Ksenofontov}, L.~T., {et~al.} 2005, \apjl, 619,
  L163

\bibitem[{{Katsuda} {et~al.}(2008){Katsuda}, {Tsunemi}, \&
  {Mori}}]{2008ApJ...678L..35K}
{Katsuda}, S., {Tsunemi}, H., \& {Mori}, K. 2008, \apjl, 678, L35

\bibitem[{{Kelner} {et~al.}(2006){Kelner}, {Aharonian}, \&
  {Bugayov}}]{2006PhRvD..74c4018K}
{Kelner}, S.~R., {Aharonian}, F.~A., \& {Bugayov}, V.~V. 2006, \prd, 74, 034018

\bibitem[{{Kishishita} {et~al.}(2013){Kishishita}, {Hiraga}, \&
  {Uchiyama}}]{2013A&A...551A.132K}
{Kishishita}, T., {Hiraga}, J., \& {Uchiyama}, Y. 2013, \aap, 551, A132

\bibitem[{{Lande} {et~al.}(2012){Lande}, {Ackermann}, {Allafort}, {Ballet},
  {Bechtol}, {Burnett}, {Cohen-Tanugi}, {Drlica-Wagner}, {Funk}, {Giordano},
  {Grondin}, {Kerr}, \& {Lemoine-Goumard}}]{paper:fermi_ext_sources}
{Lande}, J., {Ackermann}, M., {Allafort}, A., {et~al.} 2012, \apj, 756, 5

\bibitem[{{Lee} {et~al.}(2013){Lee}, {Slane}, {Ellison}, {Nagataki}, \&
  {Patnaude}}]{2013ApJ...767...20L}
{Lee}, S.-H., {Slane}, P.~O., {Ellison}, D.~C., {Nagataki}, S., \& {Patnaude},
  D.~J. 2013, \apj, 767, 20

\bibitem[{{Meyer} {et~al.}(2010){Meyer}, {Horns}, \&
  {Zechlin}}]{paper:crab_meyer}
{Meyer}, M., {Horns}, D., \& {Zechlin}, H.-S. 2010, \aap, 523, A2

\bibitem[{{Nolan} {et~al.}(2012){Nolan}, {Abdo}, {Ackermann}, {Ajello},
  {Allafort}, {Antolini}, {Atwood}, {Axelsson}, {Baldini}, {Ballet}, \&
  et~al.}]{paper:fermi_2fgl_cat}
{Nolan}, P.~L., {Abdo}, A.~A., {Ackermann}, M., {et~al.} 2012, \apjs, 199, 31

\bibitem[{{Obergaulinger} {et~al.}(2014){Obergaulinger}, {Iyudin},
  {M{\"u}ller}, \& {Smoot}}]{2014MNRAS.437..976O}
{Obergaulinger}, M., {Iyudin}, A.~F., {M{\"u}ller}, E., \& {Smoot}, G.~F. 2014,
  \mnras, 437, 976

\bibitem[{{Ohm} {et~al.}(2009){Ohm}, {van Eldik}, \&
  {Egberts}}]{paper:hess_tmva}
{Ohm}, S., {van Eldik}, C., \& {Egberts}, K. 2009, Astroparticle Physics, 31,
  383

\bibitem[{{Parizot} {et~al.}(2006){Parizot}, {Marcowith}, {Ballet}, \&
  {Gallant}}]{parizot06}
{Parizot}, E., {Marcowith}, A., {Ballet}, J., \& {Gallant}, Y.~A. 2006, \aap,
  453, 387

\bibitem[{{Piron} {et~al.}(2001){Piron}, {Djannati-Atai}, {Punch}, {Tavernet},
  {Barrau}, {Bazer-Bachi}, {Chounet}, {Debiais}, {Degrange}, {Dezalay},
  {Espigat}, {Fabre}, {Fleury}, {Fontaine}, {Goret}, {Gouiffes}, {Khelifi},
  {Malet}, {Masterson}, {Mohanty}, {Nuss}, {Renault}, {Rivoal}, {Rob}, \&
  {Vorobiov}}]{paper:ff_spectrum_piron}
{Piron}, F., {Djannati-Atai}, A., {Punch}, M., {et~al.} 2001, \aap, 374, 895

\bibitem[{{Pohl} {et~al.}(2005){Pohl}, {Yan}, \&
  {Lazarian}}]{2005ApJ...626L.101P}
{Pohl}, M., {Yan}, H., \& {Lazarian}, A. 2005, \apjl, 626, L101

\bibitem[{{Porter} {et~al.}(2006){Porter}, {Moskalenko}, \&
  {Strong}}]{2006ApJ...648L..29P}
{Porter}, T.~A., {Moskalenko}, I.~V., \& {Strong}, A.~W. 2006, \apjl, 648, L29

\bibitem[{{Rettig} \& {Pohl}(2012)}]{2012A&A...545A..47R}
{Rettig}, R. \& {Pohl}, M. 2012, \aap, 545, A47

\bibitem[{{Reynolds}(1998)}]{reynolds98}
{Reynolds}, S.~P. 1998, \apj, 493, 375

\bibitem[{{Reynoso} {et~al.}(2006){Reynoso}, {Dubner}, {Giacani}, {Johnston},
  \& {Green}}]{2006A&A...449..243R}
{Reynoso}, E.~M., {Dubner}, G., {Giacani}, E., {Johnston}, S., \& {Green},
  A.~J. 2006, \aap, 449, 243

\bibitem[{{Slane} {et~al.}(2001){Slane}, {Hughes}, {Edgar}, {Plucinsky},
  {Miyata}, {Tsunemi}, \& {Aschenbach}}]{2001ApJ...548..814S}
{Slane}, P., {Hughes}, J.~P., {Edgar}, R.~J., {et~al.} 2001, \apj, 548, 814

\bibitem[{{Stupar} {et~al.}(2005){Stupar}, {Filipovi{\'c}}, {Jones}, \&
  {Parker}}]{proc:vjr_radio_xrays}
{Stupar}, M., {Filipovi{\'c}}, M.~D., {Jones}, P.~A., \& {Parker}, Q.~A. 2005,
  Advances in Space Research, 35, 1047

\bibitem[{{Takeda} {et~al.}(2016){Takeda}, {Bamba}, {Terada}, {Tashiro},
  {Katsuda}, {Yamazaki}, {Ohira}, \& {Iwakiri}}]{2016PASJ...68S..10T}
{Takeda}, S., {Bamba}, A., {Terada}, Y., {et~al.} 2016, \pasj, 68, S10

\bibitem[{{Tanaka} {et~al.}(2011){Tanaka}, {Allafort}, {Ballet}, {Funk},
  {Giordano}, {Hewitt}, {Lemoine-Goumard}, {Tajima}, {Tibolla}, \&
  {Uchiyama}}]{paper:vjr_fermi}
{Tanaka}, T., {Allafort}, A., {Ballet}, J., {et~al.} 2011, \apjl, 740, L51

\bibitem[{{Telezhinsky}(2009)}]{2009APh....31..431T}
{Telezhinsky}, I. 2009, Astroparticle Physics, 31, 431

\bibitem[{Wilks(1938)}]{paper:wilks_theorem}
Wilks, S.~S. 1938, The Annals of Mathematical Statistics, 9, 60

\bibitem[{{Wright} {et~al.}(1994){Wright}, {Griffith}, {Burke}, \&
  {Ekers}}]{paper:radio_pmn_survey}
{Wright}, A.~E., {Griffith}, M.~R., {Burke}, B.~F., \& {Ekers}, R.~D. 1994,
  \apjs, 91, 111

\end{thebibliography}
%

\pagebreak[4] 

\begin{appendix}

\section{Region definitions and spectral points}

\begin{table*}[!htb]
  \centering
  \begin{tabular}{l|cccccc|c}
 \hline
  region         & RA            & Dec             & $R_1$         & $R_2$           & $\phi_1$      & $\phi_2$      & area            \\
                 & [$\UNITSwoS{deg}$] & [$\UNITSwoS{deg}$] & [$\UNITSwoS{deg}$] & [$\UNITSwoS{deg}$]   & [$\UNITSwoS{deg}$] & [$\UNITSwoS{deg}$] & [$\UNITSwoS{deg^2}$] \\
  \hline
  \hline
  whole SNR      & 133.00        & $-46.37$        & 0.0           & 1.0             &   0           & 360           & 3.14            \\
  \hline
  NW rim         & 133.10        & $-46.30$        & 0.6           & 1.0             & 220           &  40           & 1.01            \\
  \hline
  0              & 133.00        & $-46.37$        & 0.0           & 0.6             &   0           & 360           & 1.13            \\
  1              & 133.00        & $-46.37$        & 0.6           & 1.0             &   0           &  60           & 0.335           \\
  2              & 133.00        & $-46.37$        & 0.6           & 1.0             &  60           & 120           & 0.335           \\
  3              & 133.00        & $-46.37$        & 0.6           & 1.0             & 120           & 180           & 0.335           \\
  4              & 133.00        & $-46.37$        & 0.6           & 1.0             & 180           & 240           & 0.335           \\
  5              & 133.00        & $-46.37$        & 0.6           & 1.0             & 240           & 300           & 0.335           \\
  6              & 133.00        & $-46.37$        & 0.6           & 1.0             & 300           & 360           & 0.335           \\
  \hline
  A (point-like) & 133.90        & $-46.74$        & 0.0           & $\sqrt{0.0125}$ &   0           & 360           & 0.0393          \\
  B              & 133.85        & $-46.65$        & 0.0           & 0.3             &   0           & 360           & 0.283           \\
  C              & 133.50        & $-46.75$        & 0.0           & 0.6             &   0           & 360           & 1.13            \\
  D (point-like) & 133.25        & $-47.15$        & 0.0           & $\sqrt{0.0125}$ &   0           & 360           & 0.0393          \\
  B$'$           & \multicolumn{6}{|c|} {$\mathrm{B}' = \mathrm{B}\setminus\mathrm{A}$}                              & 0.243           \\
  C$'$           & \multicolumn{6}{|c|} {$\mathrm{C}' = \mathrm{C}\setminus\mathrm{B}\setminus\mathrm{D}$}           & 0.809           \\
\hline
  \end{tabular}
  \caption[Region definitions]{Region definitions used for the analysis of \velajr. The regions are specified with the following parameters of an annular region: the center coordinates in (RA, Dec, J2000), the inner and outer radii ($R_1$ and $R_2$, respectively) and the starting and ending polar angles ($\phi_1$ and $\phi_2$, respectively). The polar angles are defined in a similar way as the azimuthal angles in Fig.~\ref{fig:az_prof}: starting at the 12 o'clock position (the north) and increasing counterclockwise. The complex regions B$'$ and C$'$ are defined in terms of simple regions. In addition, the area of the sky covered by the corresponding region is also given.}
  \label{tab:reg_def}
\end{table*}

\begin{table*}[!htb]
  \centering
  \begin{tabular}{cccc}
  \hline
   $E$ [$\UNITSwoS{TeV}$] & $\mathrm{d}\Phi/\mathrm{d}E$ [$\UNITSwoS{cm^{-2} s^{-1} TeV^{-1}}$] & excess & sign \\
  \hline
    0.407 & $\Big(1.23 ^{+ 0.28\stat} _{- 0.28\stat} \pm 0.42\syst\Big) \times 10^{-10}$ &  329 & $ 4.4\sigma$ \\
    0.571 & $\Big(8.32 ^{+ 0.76\stat} _{- 0.76\stat} \pm 2.23\syst\Big) \times 10^{-11}$ & 1240 & $11  \sigma$ \\
    0.826 & $\Big(3.85 ^{+ 0.31\stat} _{- 0.31\stat} \pm 0.95\syst\Big) \times 10^{-11}$ & 1230 & $13  \sigma$ \\
    1.21  & $\Big(1.85 ^{+ 0.14\stat} _{- 0.14\stat} \pm 0.41\syst\Big) \times 10^{-11}$ & 1030 & $14  \sigma$ \\
    1.78  & $\Big(9.06 ^{+ 0.63\stat} _{- 0.62\stat} \pm 1.90\syst\Big) \times 10^{-12}$ &  880 & $15  \sigma$ \\
    2.60  & $\Big(4.05 ^{+ 0.31\stat} _{- 0.31\stat} \pm 0.85\syst\Big) \times 10^{-12}$ &  666 & $13  \sigma$ \\
    3.80  & $\Big(1.50 ^{+ 0.15\stat} _{- 0.15\stat} \pm 0.33\syst\Big) \times 10^{-12}$ &  391 & $10  \sigma$ \\
    5.55  & $\Big(5.69 ^{+ 0.65\stat} _{- 0.65\stat} \pm 1.31\syst\Big) \times 10^{-13}$ &  216 & $ 8.9\sigma$ \\
    8.13  & $\Big(2.37 ^{+ 0.33\stat} _{- 0.32\stat} \pm 0.53\syst\Big) \times 10^{-13}$ &  136 & $ 7.4\sigma$ \\
   11.9   & $\Big(6.47 ^{+ 1.54\stat} _{- 1.51\stat} \pm 1.41\syst\Big) \times 10^{-14}$ &   59 & $ 4.3\sigma$ \\
   17.3   & $\Big(1.35 ^{+ 0.69\stat} _{- 0.68\stat} \pm 0.33\syst\Big) \times 10^{-14}$ &   20 & $ 2.0\sigma$ \\
   25.0   & $\Big(3.33 ^{+ 2.74\stat} _{- 2.69\stat} \pm 1.20\syst\Big) \times 10^{-15}$ &    8 & $ 1.2\sigma$ \\
\hline
  \end{tabular}
  \caption[\HESS\ spectrum points]{\HESS\ flux points for the spectrum of \velajr. The table shows the energy $E$, the differential flux $\mathrm{d}\Phi/\mathrm{d}E$ with $1\sigma$ statistical and systematic uncertainties, the excess and the significance of the point in number of Gaussian standard deviations $\sigma$.}
  \label{tab:spectrum_points}
\end{table*}

\end{appendix}
%

\listofobjects 
%

\end{document}